\documentclass[11pt,preprint]{aastex}
\usepackage{color}
\usepackage{natbib}
\shorttitle{Near-Infrared Spectroscopy of Low Mass X-ray Binaries:  Accretion Disk Contamination and Mass Determination}
\shortauthors{Khargharia et al.}

\begin{document}

%% LaTeX will automatically break titles if they run longer than
%% one line. However, you may use \\ to force a line break if
%% you desire.

\title{Near-Infrared Spectroscopy of Low Mass X-ray Binaries : \\ Accretion Disk Contamination and Compact Object Mass Determination in\\ V404 Cyg and Cen X-4}

%% Use \author, \affil, and the \and command to format
%% author and affiliation information.
%% Note that \email has replaced the old \authoremail command
%% from AASTeX v4.0. You can use \email to mark an email address
%% anywhere in the paper, not just in the front matter.
%% As in the title, use \\ to force line breaks.

\author{Juthika Khargharia\altaffilmark{1}, Cynthia S. Froning\altaffilmark{1,2}}
\affil{Astrophysical and Planetary Sciences}
\affil{University of Colorado, 391, UCB \\ Boulder, CO, 80309}
\email{juthika.khargharia@colorado.edu}

\and

\author{Edward L. Robinson}
\affil{Department of Astronomy}
\affil{University of Texas at Austin, Austin, TX 78712}
\email{elr@astro.as.utexas.edu}

%% Notice that each of these authors has alternate affiliations, which
%% are identified by the \altaffilmark after each name.  Specify alternate
%% affiliation information with \altaffiltext, with one command per each
%% affiliation.

\altaffiltext{1}{Center for Astrophysics and Space Astronomy, University of Colorado, 593, UCB, Boulder, CO 80309-0593}
\altaffiltext{2}{Visiting Astronomer at the Infrared Telescope Facility, which is operated by the University of Hawai under Cooperative Agreement no. NCC 5-538 with the National Aeronautics and Space Administration, Science Mission Directorate, Planetary Astronomy Program.}

%% Mark off your abstract in the ``abstract'' environment. In the manuscript
%% style, abstract will output a Received/Accepted line after the
%% title and affiliation information. No date will appear since the author
%% does not have this information. The dates will be filled in by the
%% editorial office after submission.

\begin{abstract}
We present near-infrared (NIR) broadband (0.80--2.42 $\mu$m) spectroscopy of two low mass X-ray binaries: V404 Cyg and Cen X-4. One important parameter required in the determination of the mass of the compact objects in these systems is the binary inclination. We can determine the inclination by modeling the ellipsoidal modulations of the Roche-lobe filling donor star, but the contamination of the donor star light from other components of the binary, particularly the accretion disk, must be taken into account. To this end, we determined the donor star contribution to the infrared flux by comparing the spectra of V404 Cyg and Cen X-4 to those of various field K-stars of known spectral type. For V404 Cyg, we determined that the donor star has a spectral type of K3 III. We determined the fractional donor contribution to the NIR flux  in the H- and K-bands as $0.98 \pm .05$ and $0.97 \pm .09$, respectively. We remodeled the H-band light curve from \citet{sanwal1996} after correcting for the donor star contribution to obtain a new value for the binary inclination. From this, we determined the mass of the black hole in V404 Cyg to be $M_{BH}= 9.0^{+.2}_{-.6}M_{\odot}$. We performed the same spectral analysis for Cen X-4 and found the spectral type of the donor star to be in the range K5 -- M1V. The donor star contribution in Cen X-4 is $0.94\pm.14$ in the H-band while in the K-band, the accretion disk can contribute up to 10\% of the infrared flux. We remodeled the H-band light curve from \citet{shahbaz1993}, again correcting for the fractional contribution of the donor star to obtain the inclination. From this, we determined the mass of the neutron star as $M_{NS}= 1.5^{+.1}_{-.4}M_{\odot}$. However, the masses obtained for both systems should be viewed with some caution since contemporaneous light curve and spectral data are required to obtain definitive masses.
\end{abstract}

%% Keywords should appear after the \end{abstract} command. The uncommented
%% example has been keyed in ApJ style. See the instructions to authors
%% for the journal to which you are submitting your paper to determine
%% what keyword punctuation is appropriate.


%% From the front matter, we move on to the body of the paper.
%% In the first two sections, notice the use of the natbib \citep
%% and \citet commands to identify citations.  The citations are
%% tied to the reference list via symbolic KEYs. The KEY corresponds
%% to the KEY in the \bibitem in the reference list below. We have
%% chosen the first three characters of the first author's name plus
%% the last two numeral of the year of publication as our KEY for
%% each reference.

%% Authors who wish to have the most important objects in their paper
%% linked in the electronic edition to a data center may do so by tagging
%% their objects with \objectname{} or \object{}.  Each macro takes the
%% object name as its required argument. The optional, square-bracket 
%% argument should be used in cases where the data center identification
%% differs from what is to be printed in the paper.  The text appearing 
%% in curly braces is what will appear in print in the published paper. 
%% If the object name is recognized by the data centers, it will be linked
%% in the electronic edition to the object data available at the data centers  
%%
%% Note that for sources with brackets in their names, e.g. [WEG2004] 14h-090,
%% the brackets must be escaped with backslashes when used in the first
%% square-bracket argument, for instance, \object[\[WEG2004\] 14h-090]{90}).
%%  Otherwise, LaTeX will issue an error. 

\newpage

\section{Introduction}

V404 Cyg and Cen X-4 belong to the class of low-mass X-ray binaries (LMXBs) in which mass transfer from a late type star to a black hole (BH) or a neutron star (NS) takes place via an accretion disk. By determining the mass of the compact object, we can constrain models for the formation and evolution of black holes and neutron stars in close binary systems \citep{brown98, fryer2001, nelemans2001}. Of the 16 BH systems with BH masses determined to accuracies of 5--30\%,  most of the masses range from 4 -- 16 M$_{\odot}$ with only one system, J0422+32, with a BH mass $<$5 M$_\odot$ \citep{charlescoe06, casares2006, gelino2003}.  This result is not in agreement with the predictions of BH mass distributions from theoretical modeling which predict a continuous decreasing mass distribution from neutron star masses up to 10 -- 20$M_{\odot}$ \citep{fryer2001, belczynski08}. However, there currently are not enough systems with precise mass determinations to allow us to distinguish between observational biases and/or theoretical limitations as the source of the disparity.  To increase the number of LMXBs with precise compact object mass determinations, we have undertaken NIR spectroscopy of several systems to determine the relative contributions of the donor star and the accretion disk to the NIR light curves, from which accurate binary inclinations and, in turn, stellar masses are derived.  In this paper, we present results for two LMXBs, the BH binary V404 Cyg and the NS binary Cen X-4.

V404 Cyg went into outburst on May 22, 1989 and was subsequently discovered with the all-sky monitor aboard the Ginga satellite. A high X-ray luminosity of $\approx 10^{38}$ ergs/sec suggested that the system was a close binary system with a black hole or a neutron star as the accretor \citep{marsden89}. During the outburst, the visual magnitude of V404 Cyg increased from V=18.3 to V=11.6. One year after its discovery, it had fallen back to V=18.0 \citep{wagner92}. Since then, V404 Cyg has been studied in quiescence at optical and infrared wavelengths, establishing most of the system parameters. \citet{casares92} established the mass function, the minimum mass of the compact object, as $f(M) = 6.26 \pm 0.31 M_{\odot}$ which was later refined by \citet{casares94} to be $f(M) = 6.08 \pm 0.06 M_{\odot}$. They determined the donor star to be of spectral type K0IV and obtained the binary mass ratio from the rotational broadening of the absorption lines in the donor star spectrum.  The ellipsoidal modulation of the donor star has been modeled in the NIR to determine the binary inclination and the black hole mass: in the K-band by \citet{shahbaz1994} and in H by \citet{sanwal1996}.  The resulting black hole mass determinations were 12 M$_{\odot}$ and $<$12.5 M$_{\odot}$, respectively, where one of the key differences between their results was that \citet{shahbaz1994}  assumed that the donor star was the only source of flux in the NIR, while \citet{sanwal1996} found evidence for accretion disk contamination.  \citet{shahbaz1996} tested this assumption by performing NIR spectroscopy of V404 Cyg.  By comparing the NIR spectrum to scaled template spectra from spectral type standard stars, they determined that the donor star contributes 100$\pm$11 \% of the K-band flux.  However, for this analysis they relied on fits to the $^{12}$CO bands in K, which in other LMXBs (and cataclysmic variables) have been found to be affected by metallicity variations in the carbon abundance \citep{froning2007}.  As a result, a broader spectral analysis that examines multiple atomic and molecular species is required to firmly tie down the donor 
star contribution in the NIR.

Cen X-4 was discovered on May 23, 1969 by the ${\it Vela}$ ${\it 5B}$ satellite \citep{Conner69} when it went into X-ray outburst. During its second outburst in 1979, it increased in visual magnitude by $\sim$6 magnitudes from its pre-outburst phase \citep{canizares80}. \citet{matsuoka80} identified the mass accreting star as a neutron star due to the fact that it displayed a Type I burst. \citet{shahbaz1993} modeled the H-band light curve to obtain a binary inclination of  31$^\circ$--54$^\circ$ and the mass of the neutron star as 0.5--2.1M$_{\odot}$. In quiescence, \citet{torres2002} performed high-resolution optical spectroscopy on Cen X-4 and determined the mass of the compact object as $0.49M_{\odot} < M < 2.49 M_{\odot}$. They determined the mass function to be $f(M) = 0.220 \pm 0.005M_{\odot}$, which was later refined by \citet{davanzo05} to $ f(M) = 0.201 \pm 0.004M_{\odot}$. The system parameters were again refined by \citet{casares2007} and much tighter constraints were imposed on the radial velocity semi-amplitude, $K_2=144.6\pm.3$ km s$^{-1}$, the orbital period, $P=.6290522\pm.0000004$ d and the binary mass ratio, $q=.20\pm.03$. The mass function adopted for this paper is calculated from the $P$ and $K_2$ values of \citet{casares2007}, which is $ f(M) = 0.197 \pm 0.001M_{\odot}$.  However, as with V404 Cyg, in all the above cases, it was assumed that the donor star was the only NIR flux source, an assumption that has not been tested in Cen X-4.

In this paper, we present broadband (0.8--2.42 $\mu$m) NIR spectroscopy of V404 Cyg and Cen X-4 obtained with SpeX on the IRTF.  By comparing the spectra to field stars of known spectral type,  we determine the donor star contribution to the infrared flux for both systems. We use these results to remodel previously obtained light curve data to determine the binary inclinations and to derive precise compact star masses for both targets.

\section{Observations}

V404 Cyg and Cen X-4 were observed on June 7--9, 2007 with SpeX at the NASA Infrared Telescope Facility \citep{rayner2003}. The weather was clear with good seeing conditions ($\le0.7\arcsec$).  All observations were made using the Short XD (SXD) mode through the $0.5\arcsec$ slit (R=1200), which covers the entire wavelength range, 0.8--2.4$\mu$m,  simultaneously in 6 orders. Due to the good seeing conditions and the narrow slit used, the observed spectrum is uncontaminated by an accidental companion star at $1.5\arcsec$ from V404 Cyg. An A0V star was observed hourly for the purpose of telluric correction.  Observations were taken in ABBA pairs in two positions along the slit with a 300 sec exposure time at each position. The slit was oriented along the parallactic angle throughout the observations, limiting the relative flux calibration uncertainties to $<2\%$ \citep{vacca2003}. Table 1 lists the observations along with the corresponding cumulative exposure times and orbital phases observed.

Data reduction consisted of calibration (sky subtraction, flat-fielding, wavelength calibration), optimal spectral extraction, and telluric correction, all of which were carried out using the idl based package Spextool \citep{cushing04}. Each exposure was shifted to the rest frame of the donor star, using the orbital ephemeris from \citet{casares94} for V404 Cyg and from \citet{casares2007}  for Cen X-4.  The individual shifted exposures were median-combined and all 6 orders were merged to obtain the time-averaged spectrum. Using the magnitudes of the A0V star, the telluric correction tool, xtellcor, also flux calibrates the spectra. Given our excellent observing conditions, the absolute fluxes should be fairly accurate, although we have not made any attempt to quantify the accuracy of those values, as they are not necessary for our analysis.

 Figure 1 shows the time-averaged spectrum of V404 Cyg. %Accurate orbital phases are needed when co-adding the spectra. We propagated the errors in the orbital ephemeris from \citet{casares94} and found that the phase error at the time of acquisition of the data translated to a wavelength shift (for the time-averaged spectra) that was small compared to our 250 km s$^{-1}$ spectral resolution}. 
The spectrum has been boxcar smoothed by 3 pixels, corresponding to one resolution element. Statistical errors were propagated through all the data reduction and processing steps. The spectrum was corrected for interstellar reddening using a dereddening value of  E(B-V)= 1.303 \citep{hynes2009, miller2009}. The spectrum in red in Figure 1 is the dereddened spectrum of V404 Cyg. Shown in Figure 2, 3, and 4 are the J-, H-, and K-band spectra.  We have labeled the prominent spectral features using line identifications from \citet{wallace97}, \citet{wallace00}, \citet{kleinmann1986}, \citet{meyer98}, \citet{kirk1993}, and the Atomic line list\footnote{http://www.pa.uky.edu/~peter/atomic}. Figure 5 shows the time-averaged spectrum of Cen X-4.  It has been boxcar smoothed by 3 pixels and dereddened assuming E(B-V)= 0.1 \citep{blair84}. Figures 6, 7, and 8 shows the J-, H-,  and K-band spectra of Cen X-4. We estimated the S/N in the time-averaged spectra of V404 Cyg and Cen X-4 by fitting straight lines through various continuum-dominated regions  and measuring the scatter around these fits. For Cen X-4 we obtained a S/N $\approx$ 15--20 in J and K, and $\approx$ 25--30 in H. As the donor in V404 Cyg is likely a giant or sub-giant, there are fewer regions of pure continuum. Therefore, we obtained a lower limit to the S/N of $\approx$ 100 in H and K, and $\approx$ 50 in J. 

\section{Analysis}

\subsection{V404 Cyg}

The spectrum of V404 Cyg shows prominent narrow absorption lines of neutral metals including \ion{Al}{1}, \ion{Na}{1}, \ion{K}{1}, \ion{Mg}{1}, \ion{Fe}{1}, \ion{Si}{1}, \ion{Mg}{1}, \ion{Ti}{1} and \ion{Ca}{1} . The H-and K-bands also show molecular absorption bands of $^{12}$CO and $^{13}$CO. The absorption lines originate in the donor star. In contrast to many other X-ray binaries, V404 Cyg does not show any strong emission lines in its infrared spectrum. \citet{shahbaz1996} performed K-band spectroscopy of V404 Cyg and found no strong emission features either, implying that the accretion disk contribution in V404 Cyg in quiescence is probably very low at all times. In the K-band, we see both $^{12}$CO and $^{13}$CO features which are a strong indication of an evolved donor star.  In this paper, we make use of the entire broadband infrared spectrum to ascertain the spectral type of the donor star and then obtain a precise value for the relative contributions of the donor star and the accretion disk. 

First, we compared the broadband spectrum of V404 Cyg to field stars of different spectral types. These spectra were obtained from the IRTF Spectral Library\footnote{http://irtfweb.ifa.hawaii.edu/~spex/WebLibrary/index.html} which consists  of SpeX SXD spectra of stars of known spectral type \citep{rayner2009}. The template stars used in our analysis have roughly solar metallicities and similar S/N to each other. We only used non-variable stars for our analysis. Earlier work done on the spectral type of the donor star in V404 Cyg suggested a K0IV spectral type \citep{casares94}. In the IRTF spectral library, the closest available spectral type to a K0 IV is a K1 IV.  In Figure 9, we compare the spectrum of a K1 IV (HD165438) star to the spectrum of V404 Cyg, normalizing the spectrum to that of V404 Cyg just blueward of 2.29$\mu$m. It is evident that the CO bands in the template spectrum are much weaker compared to those in V404 Cyg. Also, if the template is normalized to V404 Cyg in K, its flux exceeds the J-band flux in V404 Cyg flux by $>30\%$, indicating that a K1 IV spectrum is too blue to match the donor star spectrum in V404 Cyg, unless there is another NIR flux source diluting its contribution.  

\citet{hynes2009} did a multiwavelength study of the spectral energy distribution (SED) of V404 Cyg in quiescence and found no substantial contribution from the accretion disk from near-UV to near-IR. They also found that for the donor, the spectrum of a K0 III star was in agreement with the data from UV to near-IR. At the same time, their multiwavelength SED comparison with a G8 III and a K2 III star could not be distinguished from a K0 III. We find that a K2 III (HD 132935) spectrum exceeds most of the J-band (.8 -- 1.1$\mu$ m) flux by $>$7\% when normalized in K, indicating that a K2 III donor star spectral type would require a diluting flux source in the NIR. Also, the CO bands in a K2 III appear too weak to match that of V404 Cyg in the K-band. In Figure 10, we compare a K3 III (HD 221246) spectrum to that of V404 Cyg.  In this case, the match is much closer, both in the broadband shape and spectral fluxes and in the relative depths of the CO bands. When normalized to a 100\% in K, we find that the K3 III template exceeds the J-band, near the band edge (.8 -- .9 $\mu$ m) by at most 5\%. On the other hand, the K4 III (HD 207991) spectrum falls below the V404 Cyg spectrum in J by 20\%, so that if that spectral type is correct, the donor star may dominate the NIR flux in H and K, with an increasing disk contribution at bluer wavelenghs. In conclusion, both the spectral shape and the strength of the CO bands in V404 Cyg are inconsistent with a K1 IV spectral type for the donor star and indicate that a more evolved spectral type is better match to the donor star. In the next section, we calculate the equivalent widths of various lines of V404 Cyg and compare that to the same features in field star populations to determine the spectral type more precisely. 

\subsubsection{Equivalent width analysis}

\citet{fs2000}, hereafter FS, used the equivalent width (EW) ratios of various NIR absorption lines in a population of field giants and supergiants (also a few dwarfs) as indices to determine stellar effective temperature and gravity as well as evidence of dilution from non-stellar sources. In Table 2, we list the equivalent widths for V404 Cyg for the diagnostic features in FS. The error bars represent uncertainties in continuum placement only.  By comparing our EWs to those in Figure 5 of FS, we obtain a donor effective temperature range of 4200--4600 K, implying a spectral type between K2 III and K4 III, assuming that the donor star is the only NIR flux source.  FS also demonstrated how the EW ratios of certain lines would change if there was a source of diluting continuum emission. In their Figure 8 they plotted dilution-free indices (ratios of atomic/molecular EWs) and directional indicators to show how the indices would change as the contribution from a diluting source increases. A comparison of the same EW ratios in V404 Cyg to their indices shows that the V404 Cyg ratios fall on their plots for the non-diluted sources, suggesting that the amount of dilution in V404 Cyg in both H and K is negligible. 

\subsubsection{Field star fits to V404 Cyg}

To obtain the fraction of NIR continuum emission originating from the donor star, $f$, we compared the spectrum of V404 Cyg with those of field stars ranging from K2 III to K4 III spectral types. The spectrum of the template star was convolved with a Gaussian of 38.8 km s$^{-1}$ FWHM in order to take into account the rotational velocity of the donor star in V404 Cyg (but we note that this correction was negligible given the 250 km s$^{-1}$ resolution of our spectra). Before fitting,  we normalized the spectra of both V404 Cyg and the template by fitting cubic splines along the local continuum surrounding the lines under analysis. The continuum points were picked by eye. 
%Since the NIR spectra of giants have few regions of pure continuum and the S/N of V404 Cyg spectrum was relatively high, our hand-picked points were biased toward the upper envelope of the spectra.  

We determined the donor fraction as follows:  after normalizing  both the target and the template spectra, we multiplied the latter by a fraction, $f$, which was incremented from 0 to 1 in steps of 0.01. The result was subtracted from the spectrum of V404 Cyg to obtain the residual. We then minimized the chi-square,  $\chi^{2}_{\nu}$, between the residual and the mean of the residual to obtain the donor fraction that gave us the best fit. In cases where the template had weaker absorption features compared to V404 Cyg, we let the donor fraction, $f$, take on values greater than 1. Even though this represents an unphysical scenario, it is useful as a guide to constrain the correct spectral type of the donor star by calculating the extent of weakness of the absorption feature in a particular template. 
%We fit different spectral types to the normalized spectrum of V404 Cyg for various wavelength ranges. 
%Our choice for the template star that best represented the donor in V404 Cyg was based on the lowest reduced chi-square value obtained among the different spectral types for particular absorption features and also on the donor fraction ($f$) itself, in cases where $f>1$.  
For the fitting procedure, we selected only the strongest absorption features and fit absorption features for individual species separately.  
%(Although, as FS noted, even strong transitions in the NIR spectra of late-type stars are typically blend from several species.) 
There were cases where absorption features of two different species occurred very close together in wavelength. If we were not able to separate them and identify the continuum around those absorption features, we did not include them in our analysis. 

%In the K-band,  the \ion{Mg}{1} + \ion{Al}{1}  feature in the wavelength region $\lambda$2.108 -- 2.110 $\mu$m falls under that category. The \ion{Fe}{1} features were also excluded from the analysis since they were not as strong as \ion{Na}{1} or \ion{Ca}{1} lines. In the 1.9 -- 2.0$\mu$m region there are several \ion{Ca}{1} features from which we selected the most prominent ones (1.97-- 2.00$\mu$m) for the fit. We excluded the features that lie very near the band edge where any residual from telluric correction would most affect the fit results.  In the J-band, for the region 0.8 -- 1.1$\mu$m, we considered only the strongest features which include \ion{Mg}{1} and \ion{Ca}{1} at $\lambda$0.856$\mu$m and 0.867$\mu$m respectively. We also fit several lines from 1.15 -- 1.35$\mu$m but excluded the \ion{K}{1}  line at 1.178$\mu$m, \ion{Fe}{1} + \ion{Si}{1} (1.198 -- 1.200)$\mu$m and \ion{Mg}{1} at 1.209$\mu$m. For the H-band all the identified lines in Figure 3 were taken into consideration. For the K- and the H-band, besides fitting individual prominent lines, we also fit the whole band at once. For the entire J-band (0.8 -- 1.4 $\mu$m), we were not able to normalize the entire band at once satisfactorily, given the spectral shape and the noise in this band.

In order to better account for the uncertainty associated with the data, we divided our individual exposures into two sets. They were then separately combined and the result of one divided by the other. The deviation from the mean was then determined for each spectral order separately to assign the errors. Since the J-band comprises of three orders, we obtained the mean of these to assign the error bars for the J-band. 
%For the H- and K-band, we used the corresponding deviation from the mean in orders 3 and 4 for the errors. 
We used these errors for the $\chi^2$ analysis for V404 Cyg as it was clear that the statistical error bars are too small and do not account for systematic uncertainties in the data reduction process. For the template stars we used the statistical errors provided with the template spectra, but we note that the uncertainties for V404 Cyg dominate over the error in the template stars while fitting various spectral features.  

Table 3 shows the donor fractions calculated for K2 III, K3 III and K4 III spectral types along with the $\chi^{2}_{\nu}$ values. When we compare the $\chi^{2}_{\nu}$ values for fits across the entire K and H-band at once, we obtain K3 III as the best match for the donor in V404 Cyg. However, in order to make our final choice on the donor spectral type, we also looked at all the individual absorption features selected for our analysis. 

In the K-band, we fit the CO bands together from 2.29 -- 2.42 $\mu$m and obtained the best fit with a K3 III spectral type.  There is a Na I feature near 2.34 $\mu$m included in the fit region but the fit is dominated by the prominent CO-features. Generally, the K3 III star has the lowest $\chi^{2}_{\nu}$ for fits to individual absorption features compared to the K2 III or the K4 III template stars. For the K2 III  star all the absorption features are consistently weak in the template compared to V404 Cyg (resulting in the unphysical fit, $f>1$). The K4 III template has fewer $f>1$ deviations than for the K2 III star, but some of the lines cannot be fit with $f<1$ and the $\chi^{2}_{\nu}$ values are generally larger than for a K3 III template. Even for the K3 III template, some lines give $f>1$ including the \ion{Mg}{1},  \ion{Ca}{1} and the \ion{Al}{1} feature near 2.28 $\mu$m, 1.98 $\mu$m and 2.112 -- 2.128$\mu$m, respectively. These features deviate less from $f=1$ for a K3 III star than for the other templates but may indicate systematic mismatches (in temperature, gravity, or metallicity) between V404 Cyg and the field star templates. Overall, a K3 III star provides the most consistent match to the K-band spectrum of V404 Cyg.

In the H-band, looking at the CO features, we find $f=1.24$ for the K2 III template, implying very weak $^{12}$CO features in the template compared to V404 Cyg, whereas in K3 III and K4 III, the value of $f$ is $0.98$ and $0.95$, respectively. In case of the K2 III star, the value of donor fraction is consistently $>1$ for all the individual features in the H-band also indicating that most of its absorption features are weaker than in V404 Cyg. With the K3 III and K4 III stars, we find that \ion{Mg}{1} is always weaker in the template than in V404 Cyg  (except for one feature at 1.71$\mu$m when compared against a K4 III template). This trend is also observed for \ion{Mg}{1} line in K. The same is true for the \ion{Al}{1} feature in the H-band though it is close to $f=1$ for the K3 III template. In the H-band too, we find that a K3 III star is the most consistent match to the donor star spectrum in V404 Cyg in terms of $\chi^{2}_{\nu}$ as well as the strength of the lines indicated by the donor fraction.

In the J-band too, we find that when we compare the $\chi^{2}_{\nu}$ values and also the strength of the absorption features of the template with respect to V404 Cyg for the three different spectral types,  a K3 III spectral type best matches the donor type in V404 Cyg.  From our analysis with the spectral energy distribution, equivalent width calculation of various lines and direct determination of donor fractions, we conclude that the donor star in V404 Cyg has a K3 III spectral type. We see from Table 3 that the \ion{Mg}{1} features for a K3 III spectral type has been consistently weaker in the template than in V404 Cyg (except the one feature near .856$\mu$m). This suggests an abundance mismatch between the template and V404 Cyg for this species. The same is perhaps true for  \ion{Al}{1} but there are not enough features to establish a consistent trend. For this reason, we did not include \ion{Mg}{1} in our final donor fraction calculation.

Even though there is a good qualitative match between the spectrum and the K3 III template, our $\chi^{2}$ statistics are poor, ranging from $\chi^{2}_{\nu} << 1$ for fits to individual lines in J and H to $>>1$ for fits to individual lines in H and broadband fits in H and K. We attempted to take into account the systematic uncertainties that dominate in NIR ground-based spectra by recalculating the error bars as described above, but it is clear that our error bars are still too large in J and portions of K and too small in H. Due to this, we did not rely on $\chi^{2}$ statistics to obtain our final donor fraction. Instead, we determined the donor fraction in V404 Cyg by taking an average of the best fits to multiple absorption features using a K3 III template and used the standard deviation between these line fits to represent the uncertainty on the final value of $f$. We believe this is a better tracer of systematic uncertainties in our fitting process particularly because it also takes into account the possibility of slight mismatches between the V404 Cyg and the stellar template. Using this method, we obtained a H-band donor fraction of ${\it f = 0.98 \pm .05}$ and a K-band fraction of  ${\it f = 0.97 \pm .09}$. Thus, we find no substantial NIR contribution from the accretion disk in V404 Cyg similar to \citet{hynes2009} and \citet{shahbaz1996}.

%We obtained the H-band donor fraction by averaging \ion{Si}{1} (1.584-1.595)$\mu$m, CO-bands (1.615-1.650)$\mu$m, and \ion{Al}{1} (1.670-1.678)$\mu$m and determined the donor fraction as ${\it f = 0.98 \pm .05}$. For the K-band, we averaged over \ion{Ca}{1} (1.97-1.99)$\mu$m, \ion{Al}{1} (2.112-2.128)$\mu$m, \ion{Na}{1} (2.203-2.214)$\mu$m, \ion{Ca}{1} 2.26-2.27)$\mu$m and the CO-bands (2.28-2.42)$\mu$m and obtained the K-band donor fraction as ${\it f = 0.97 \pm .09}$. 
%We did not include \ion{Mg}{1} features in our calculation of the donor fraction in both the H- and the K-band due to reasons mentioned before. 

In Figure 11, we show the the K-band spectrum of V404 Cyg overplotted with the spectrum of a K3 III star scaled by a factor of 0.97. The close qualitative match between the template and spectrum is evident. The \ion{Mg}{1} lines at 2.10 and 2.28 $\mu$m are weaker in the template spectrum than in the V404 Cyg.  Shown in Figure 12 is the spectrum of V404 Cyg in the H-band over plotted with a scaled K3 III spectrum using $f=0.98$. The \ion{Mg}{1} lines at 1.50$\mu$m, 1.58$\mu$m and 1.71$\mu$m in the template are weaker than in V404 Cyg but the other absorption features give satisfactory fits.

\subsection{Cen X-4}

The spectrum of Cen X-4 shows broad emission lines of H I and He I  and narrow absorption lines of neutral metals including transitions of \ion{Na}{1}, \ion{Mg}{1}, \ion{Al}{1}, \ion{Si}{1}, \ion{Ca}{1}, \ion{Fe}{1} and molecules including $^{12}$CO. The absorption lines originate in the cool donor star while the emission lines originate in the accretion disk. \citet{chevalier89} observed Cen X-4 in quiescence and determined the donor star to be of spectral type K5V -- K7V. They also estimated that 25-30$\%$ of the total light was contributed by the accretion disk at visual wavelengths. In the infrared, the contribution of the accretion disk to the total flux has traditionally been expected to be even less. \citet{shahbaz1993} obtained the H-band light curve of Cen X-4 and predicted the contribution from the accretion disk to be significantly less than 10$\%$. 

Our spectrum of Cen X-4 is noisier compared to that of V404 Cyg which will make a precise spectral type determination more challenging. We compared the spectrum of Cen X-4 with field dwarfs ranging from K5 -- M1 V. The normalization was done at a wavelength just blueward of the $^{12}$CO bandhead near 2.29 $\mu$m.  If the donor star in Cen X-4 is the only contributor to the infrared flux, we find that the spectral types K5 -- K7 V can be excluded. When the flux of a K5V star is normalized to that of Cen X-4 in the K-band, it exceeds the flux in the H-band by $>15\%$. This excess in flux decreases as we move to cooler stars. For example, in Figure 13 we find an excess of $\approx5\%$ in a K7 V star in the H-band and in the short J-band when normalized in K. From the comparisons, we conclude that if the donor star in Cen X-4 is the only NIR flux contributor, then it must be of spectral type later than K7 V; otherwise, some dilution must be present. For later spectral types, the flux of the donor star falls below that of Cen X-4 in the J-band and matches the H-band, as seen for an M0 V star in Figure 14. We also find $\approx1\%$ (less than the error in the K-band) excess in the shorter K-band in all spectral types including an M0 V, indicating that some dilution, if quite small, may be needed for later spectral types as well.

\subsubsection{Equivalent width analysis}

The analysis in FS focuses more on giants and supergiants than dwarfs, making use of his temperature and dilution indices ideal for V404 Cyg but less so for Cen X-4. However, we calculated the equivalent widths of the absorption features used by FS. Table 4 lists the values obtained for the equivalent widths of several absorption lines in Cen X-4. Due to the lack of data on dwarfs in FS coupled with a low S/N Cen X-4 spectrum, we obtained a large scatter in the $T_{eff}$ values. We relied on just the prominent lines in Cen X-4 like the \ion{Ca}{1} line at 2.26$\mu$m, the $^{12}$CO bandhead at 2.29$\mu$m, the \ion{Na}{1} line at 2.2$\mu$m, and the $^{12}$CO bandhead at 1.62$\mu$m to constrain the spectral range of the donor. For \ion{Ca}{1} at 2.26$\mu$m in FS, there are no data on dwarfs for the EW we obtained. With the remaining lines, we find K7 V -- M1 V as the spectral range of the donor star in Cen X-4 in the event of no dilution.

\citet{ ali1995} have performed a similar analysis as FS but with emphasis on dwarfs ranging from F3 -- M6 V. We calculated the equivalent widths for \ion{Ca}{1}, \ion{Mg}{1}, \ion{Na}{1} and the $^{12}$CO (2-0) band using the integration limits defined in their paper, which we show in Table 5.  (Note that the FS and \citet{ali1995} \ EWs are typically different for the same spectral features because of different choices in the integration limits over which the EWs were calculated.)  We compared the EWs obtained for Cen X-4 with Figures 3 and 4 of \citet{ali1995}. Again, if we use only the prominent lines 
%like \ion{Ca}{1}, \ion{Na}{1} and $^{12}$CO band, 
we find that the EW from \ion{Na}{1} points to a spectral type K5 -- K7 V and that \ion{Ca}{1}  predicts a spectral type later than K7 V. The EW of the $^{12}$CO bandhead at 2.29$\mu$m gives inconclusive results as it falls within the flat portion of their data as well as \ion{Mg}{1}.
%(see Figure 4 in \citet{ali1995}) which points to a $T_{eff}$= 3500 -- 5000 K. 
\citet{ali1995} also noted that the sensitivity of CO to gravity makes it less useful as a temperature diagnostic. %\ion{Mg}{1} is also not a useful temperature indicator as its EW falls near the top flat region in Figure 3 of \citet{ali1995}. 
In the absence of dilution,  we estimate the range of the spectral type of the donor in Cen X-4  as K5 -- M1 V.

Our conclusion from the EW calculation using the absorption features of \citet{ali1995} is different from the estimate of FS in that we cannot rule out K5 V as a possible donor type. At the same time, from a spectral energy distribution comparison, if a spectral type as early as K5 V or K7 V were to be a possible donor in Cen X-4, we should expect to find some amount of dilution. We will examine the full K5 V -- M1 V range to find out the best spectral type for the donor in Cen X-4. In the next section, we present the results obtained from performing a dilution analysis..

\subsubsection{Field star fits to Cen X-4}

We compared the spectrum of Cen X-4 with those of field dwarfs: K5 V (HD36003), K7 V (HD237903) and M1 V (HD19305) spectral types from the IRTF spectral library using the same procedure used in calculating the donor star fraction in V404 Cyg. The spectrum of the template star was convolved with a Gaussian of 43 km s$^{-1}$ FWHM to account for the rotational velocity of the donor in Cen X-4. 
%Again, this correction is negligible given the 250 km s$^{-1}$ resolution of our spectra. 
The same procedure for determining the donor fraction in V404 Cyg was used for Cen X-4. 
%but the lower S/N of Cen X-4.
% made it more challenging when selecting the continuum points around some of the absorption features. The normalized spectra of Cen X-4 was compared with normalized K5 -- M1 V templates to obtain the donor fraction, which was allowed to take on values, $f > 1$. Our criteria for choosing the template that best represented the donor in Cen X-4 were the same as used in V404 Cyg. 
For the fitting process, we chose only the most prominent absorption features in the spectrum of Cen X-4. The absorption features of individual species were fit separately and when such features occurred too close in wavelength such that the continuum region around an individual species could not be separated, we did not include that in our analysis. Due to the lower S/N, we fitted fewer lines than in V404 Cyg. 
%For example, we excluded the \ion{Mg}{1} + \ion{Al}{1} ( 2.108-2.110)$\mu$m and \ion{Mg}{1} (2.276-2.286)$\mu$m features in the K-band. As we approach the long K-region, the spectrum gets noisier and hence we included the only identifiable $^{12}$CO bandhead (2.295-2.305) $\mu$m in our analysis. Among the \ion{Ca}{1} features occurring near the band edge, we took into account only theprominent \ion{Ca}{1} (1.97-1.99 )$\mu$m features to eliminate any telluric correction residual from affecting the fits. Similarly, in the J-band, we only fit the features occurring in the (1.108-1.207)$\mu$m range and the \ion{Al}{1} feature occurring at (1.310-1.317) $\mu$m, where we were able to obtain the continuum points around the features to our satisfaction. In the H-band, we included all the identified features in our analysis. 

In order to better account for the systematic uncertainties, we determined the error bars in Cen X-4 by fitting straight lines through various continuum-dominated regions in the J-, H- and K-bands. 
%In the K-band, we chose the wavelengths: (2.09 - 2.10 )$\mu$m, (2.13 - 2.16 )$\mu$m, (2.21 - 2.23 )$\mu$m, (2.24 - 2.25 )$\mu$m, (2.09 - 2.10 )$\mu$m and (2.30 - 2.32 )$\mu$m. In H: (1.54 - 1.56 )$\mu$m, (1.60 - 1.61 )$\mu$m and (1.72 - 1.73 )$\mu$m and in J: (1.145 - 1.158 )$\mu$m, (1.215 - 1.225 )$\mu$m and (1.258 - 1.266 )$\mu$m. 
For Cen X-4, we find this a more reliable method for error determination than the method of dividing the data into half used for V404 Cyg. (Since the NIR spectra of giants have fewer regions of continuum without strong line absorption, this method could not be applied in the error determination of V404 Cyg.) We propagated the error from the template stars as well in our analysis although the $\chi^{2}_{\nu}$ values are dominated by the uncertainty from the Cen X-4 spectrum. 
 
Table 6 shows the best fit donor fractions along with the $\chi^{2}_{\nu}$ values obtained in Cen X-4 for various wavelength ranges with K5 V -- M0 V template stars. When we compare the $\chi^{2}_{\nu}$ values for fits across the entire K and H-band at once (after masking out the emission lines), we obtain an M0 V and a K5 V as the best matches, respectively. The entire J-band could not be normalized satisfactorily at once and hence is absent in Table 6.  We also looked at the $\chi^{2}_{\nu}$ values and the donor fractions for individual lines. In the K-band, a K5 V star has the lowest $\chi^{2}_{\nu}$ for most of the features while in the H- and the J-bands, it is an M0 V and a K5 V star, respectively. For all the features, we also examined the donor fractions for cases where $f>1$. From Table 6, it is seen that in the K-band, an M0 V satisfies $f \le1$ for most cases and a K5 V does not; whereas in the H-band, all three spectral types are almost equally good, though the \ion{Si}{1} feature (1.585 -- 1.605 $\mu$m) gives $f>1$ for all three templates. The same can be said of the J-band considering that the fractions $f=1.03$ and $f=1.06$ obtained for a K7 V and an M0 V in the \ion{Mg}{1} feature are close to $f=1$ (especially given the low S/N of Cen X-4).  From this examination, it is evident that that data will not support narrowing the spectral type of the donor star from the K5V -- M0V range under consideration.

Three features (the \ion{Na}{1} line in the K-band, the \ion{Si}{1} features in the H- and J-band and the \ion{Fe}{1} feature in the J-band) are consistently weaker in all the templates than in Cen X-4. This might represent abundance mismatch between the donor in Cen X-4 and the templates akin to the mismatch seen in the \ion{Mg}{1} feature in V404 Cyg. 
%\ion{Na}{1}, for example, gives $f=1.48, 1.25$ and $1.15$ for a K5, K7 and M0 V stars, respectively, which represents a very weak \ion{Na}{1} feature in all three templates and specifically in the red-component of the doublet. Weak features are also seen in the \ion{Si}{1} and \ion{Fe}{1} lines. 
FS has noted that the temperature and luminosity variations of \ion{Ca}{1}, \ion{Na}{1} and \ion{Fe}{1} features in the K-band are primarily governed by a blend of those species with \ion{Sc}{1}, \ion{Ti}{1} and \ion{V}{1} which could explain why it is challenging to self-consistently match templates to these features. (A similar problem was encountered in K-band fits to the spectrum of A0620-00; Froning et al. 2007.) These features cannot be eliminated from the final donor fraction calculation, however. Owing to the presence of only a couple of these species in the entire spectrum, it is not possible to make a decisive statement about abundance mismatch for specific elements. Other notable features are the $^{12}$CO features in the K-band and in the H-band, which in other X-ray binaries such as A0620-00 have been anomalously weak. The CO feature in the K-band is much weaker in the template for a K5 V and a K7 V compared to Cen X-4, but all of the spectral types give consistent results of $f \sim 0.93$ for the H-band feature. A better S/N spectrum of Cen X-4 is required to make a conclusive argument about this discrepancy especially as the S/N deteriorates sharply in the long K region where the CO bandheads occur.

%We have already shown from a comparison of the spectral energy distribution that only spectral types later than K7 V can be the sole contributor to the infrared flux in Cen X-4 without exceeding the flux in H and J (neglecting the small $\approx$1\% excess in the short K-band, given the noise in this band). Now we find that all the absorption lines in a K5 V are too weak to match those of Cen X-4 in the K-band. This is also true for a K7 V for two of the four features where $f>1$. In the H-band, where the S/N in Cen X-4 is higher, all three spectral types can fit the lines equally well. From the EW analysis and the broadband spectral energy distribution comparison with Cen X-4, an M1 V spectral type could be not be ruled out as a possible donor in Cen X-4 . Since the only M1 V star (HD 42581) from the IRTF spectral library happens to be a variable star, however, we will not include it in our final donor fraction. 

We determined the final donor fraction by averaging over all the individual selected features from Table 6 for all three of the spectral type templates from K5 -- M0 V. 
%With the fit results obtained from the \ion{Mg}{1} (1.48-1.52 $\mu$m), \ion{Mg}{1} (1.57-1.582) $\mu$m, \ion{Si}{1} (1.585-1.605), $^{12}$CO (1.615-1.65) $\mu$m, \ion{Al}{1} (1.670-1.685 $\mu$m) and \ion{Mg}{1} (1.70-1.72$\mu$m) lines, 
We obtain a mean donor fraction of $f = 0.94 \pm .14$ in the H-band 
%In the K-band, we average over the \ion{Ca}{1} (1.97-1.99) $\mu$m, \ion{Na}{1} (2.203 - 2.214) $\mu$m, \ion{Ca}{1} (2.26 - 2.27) $\mu$m and $^{12}$CO (2.295 -2.305) $\mu$m line fits to 
and obtain $f=1.09\pm.20$ in K. Thus, the accretion disk can contribute up to 10\% in the K-band. If a K5 V were to be excluded from the donor spectral type (based on its weak K-band features), we find a disk contribution of up to 17\% in K. This is consistent with the H -band fraction of 0.94, which would imply a $\sim$6 -- 10\% dilution in K (depending on the template spectral type). The H-band donor fraction contribution of $f = 0.94 \pm .14$ remains the same whether the K5V star is included or not. Since the mass of the NS in Cen X-4 will be obtained from the H-band light curve data of \citet{shahbaz1993}, the inclusion or exclusion of a K5 V star will not affect the final mass. Figure 15 and 16 shows a comparison of the spectrum of Cen X-4 in the H- and K-band respectively, after scaling a K7 V and an M0 V template by a donor fraction of .94 and 1.0 respectively.  

\section{Discussion}
 
\subsection{V404 Cyg}

\subsubsection{Accretion disk contamination and mass of the black hole in V404 Cyg}

Previous estimates have given K0 IV as the donor type in V404 Cyg \citep{casares94}. Our high S/N NIR spectrum shows strong $^{12}$CO and $^{13}$CO  features that point to a more evolved star.  By comparing the spectral energy distribution with various field stars, EW diagnostics and dilution analysis, we found that a K3 III star is the best match to the donor star of V404 Cyg with minimal accretion disk contribution. From a comparison of the spectrum of V404 Cyg with a scaled K3 III star, one can appreciate the close qualitative match in the two spectra. In the K-band, the CO bands and most metals %including transitions of \ion{Ca}{1}, \ion{Si}{1}, \ion{Al}{1}  and \ion{Fe}{1}, 
are consistent with the K3 III spectral type but \ion{Mg}{1} appears supersolar in both the H- and K-bands.  Unlike other compact binaries like A0620-00 \citep{froning2007}, XTE J1118+480 \citep{haswell2002} and some cataclysmic variables \citep{harrison2000, harrison2004}  that show weak CO bandheads and depleted carbon abundances, the CO bandheads in V404 Cyg are consistent with the solar abundance K3III template spectrum.

\citet{sanwal1996} obtained an H-band light curve of V404 Cyg. We retrieved their data from their Table 2  and remodeled the light curve taking into account the (small) dilution of the donor star flux. The light curve modeling program is the same as used for A0620-00 \citep{froning2001}. The parameters used for the model are T$_{eff}$=4300 K, a gravity darkening exponent of 0.08 (Lucy, 1967) and a limb darkening coefficient from \citet{claret95} for a $\log g$ value suitable for a K3 III star \citep{berdy94, cayrel80}. For the error bars on the light curve, we used the scatter on the reference field star observed by \citet{sanwal1996} (.02 mag) and equally weighted the points. For a mass ratio of q = 0.060$^{+.004}_{-.005}$ and $f(M) = 6.08\pm.06$ \citep{casares94}, the mass of the BH can be written  to one unknown as M$_{BH}$ = ($6.83 \pm .09$) $\sin^{-3} i$, where $i$ is the inclination of the binary. We varied the donor fraction from .93 -- 1.00 and obtained a best fit ($\chi^{2}_{\nu}$ = 3.3) to the H-band light curve data for an inclination of $i= (67^{+3}_{-1})^\circ$ where the uncertainty in the inclination is propagated from the fractional donor star contribution. Figure 17 shows the best fit light curve of V404 Cyg obtained for $q=.06$ and $i=67^\circ$. The uncertainty was dominated by the change in the donor fraction. 
%The amplitude of the ellipsoidal variations depends weakly on temperature. We found a variation of about 0.3$^\circ$, as $T_{eff}$ is varied from 4300 -- 4500 K to include the variation between a K2  III and a K4 III. To check the dependence of the inclination on limb-darkening coefficient, we varied $\log g$ from 2 -- 4 to include giant and sub-giant spectral types with solar metallicity and found it to be still insignificant ($\approx 1{^\circ}$) as compared to varying the donor fraction. We also found that the change in the inclination of the binary was negligible ($\approx 0.6^\circ$) when we varied the mass-ratio within its uncertainty as given by \citet{casares94}. 
Accordingly, we obtained the mass of the black hole in V404 Cyg as $9.0^{+.2}_{-.6}$$M_\odot$ for $q=0.06$.

It has been shown in \citet{cantrell2008,cantrell2010} that fitting ellipsoidal models to the light curves of A0620-00 in its passive state increases the binary inclination considerably and results in reduced mass of the black hole than previous studies. The variability in the light curve shape (which can change from night to night) has been attributed to the disk component. Since V404 Cyg has a high S/N, we tested the stability of the donor fraction from the spectrum obtained for each individual night and found it to be fairly constant for each night. V404 Cyg exhibits fast strong variability which is likely produced by the disk \citep{sanwal1996}. To test what effect that would have on the inclination and the black hole mass, we picked points from the lower envelope of their light curve data in Figure 1 by eye. We found a best fit for an inclination of $\approx 72^\circ$ which translates to a reduction of black hole mass by about 1M$_\odot$ which is still in agreement with \citet{sanwal1996}.

\citet{shahbaz1996} obtained a 100$\%$ contribution of the donor star to the infrared flux in the K-band which implies a binary inclination of $56^\circ$. Despite our comparable result for the minimal accretion disk contribution, their inclination does not agree with our value within the error bars. They used the light curve from \citet{shahbaz1994} to determine the inclination. \citet{sanwal1996} recalculated their K-band model light curve by equally weighing each night of data and obtained a significant improvement on the variance of the model light curve. From Figure 2 of \citet{sanwal1996}, for a mass ratio of q=.07, the absolute lower limit to the inclination corresponds to $\approx$ 60$^\circ$. They established an upper limit of 12.5M$_{\odot}$ for the mass of the black hole in V404 Cyg and our result is consistent with their limits with respect to both the inclination and the BH mass.

\subsubsection{Accretion disk contamination and Mass of the neutron star in Cen X-4}

From a comparison of the spectrum of Cen X-4 to those of various field dwarfs, we found that the donor star in Cen X-4 is probably not the only source of NIR flux. We found that in the H-band, a fraction, ${0.94 \pm 0.14}$, of the NIR flux comes from the donor star while in the K-band this fraction is $1.09 \pm.20$. Earlier estimates for the donor type in Cen X-4 include a K7 V star \citep{shahbaz1993} , a K3 V--K5V star \citep{torres2002} and a K3 V--K5 V star \citep{davanzo05}. We have estimated the donor type in Cen X-4 to lie between a K5 V and an M1 V. If the donor star is a K5 or K7V, there must be at least 5 -- 15\% K-band dilution to reconcile the NIR spectral energy distribution in Cen X-4 with the template SED.  A later type donor star of M0V or M1V requires only a few percent dilution in K to match the broadband SED. Higher signal to noise data is required to constrain the donor spectral type even further. 

We use the orbital parameters from \citet{casares2007} and determine the mass function is found to be $f(M)=.197\pm.001$. Using this value of the mass function and the binary mass ratio, $q=.20\pm.03$, the mass of the neutron star in Cen X-4 can be written upto one uknown as (0.28$\pm$.01) $\sin^{-3} i$ $M_{\odot}$.

%We use the orbital parameters from \citet{casares2007} which includes the rotational velocity for the donor star in Cen X-4,  $v_{rot} \sin i$ as 44$\pm$3 km s$^{-1}$, a much tighter constraint on the velocity semi-amplitude, $K_2$= 144.6$\pm$0.3 km s$^{-1}$ and an orbital period, $P$=0.6290522$\pm$.0000004 d.  They obtained the mass ratio, q, as 0.20$\pm$0.03. With the new parameters the mass function is found to be $f(M)=.197\pm.001$.

%\citet{torres2002} determined the rotational velocity for the donor star in Cen X-4,  $v_{rot} \sin i$ as 43$\pm$6 km s$^{-1}$ and used it to obtain the mass ratio, q as 0.17$\pm$0.06. We used $v_{rot} \sin i$ from \citet{torres2002} and the radial velocity obtained by \citet{davanzo05}. The radial velocity, $K_2 = 145.8 \pm 1.0$ km $s^{-1}$, obtained by \citet{davanzo05} also agrees with two previous estimates of \citet{cowley87} and \citet{mcclintock1990}.

For modeling purposes, we chose $T_{eff} = 4300$ K for a K7 V. We chose a value of $\log{g}$=4.5  appropriate for a K7 V star to obtain the corresponding limb darkening coefficients \citep{claret95}. We also chose a value of 0.08 for the gravity darkening coefficient assuming that the donor has a convective envelope \citep{lucy1967, sarna1989}.  

 %\citet{shahbaz1993} obtained light curve data for Cen X-4 in the H-band and modeled its ellipsoidal modulation. They determined the inclination of the binary in Cen X-4 as 31$^\circ$--54$^\circ$ with which they found the mass of the neutron star of 0.5 -- 2.1M$_{\odot}$. 
 The flux from the accretion disk will reduce the apparent amplitude of the ellipsoidal modulations which in turn will underestimate the binary inclination and overestimate the mass of the neutron star.  We determined that the contamination from the accretion disk could be up to 20\% in H. We can make an initial estimate of the mass of the neutron star in Cen X-4 by combining our dilution estimate with the H-band light curve observations of Shahbaz et al.  We read in their light curve data from their Figure 1.  Their data points include uncertainties in both magnitude and orbital phase, which was derived from how they binned their data (rescaling their errors in each phase bin to get $\chi^{2}_{\nu}$ = 1 and then averaging the points).  We could not read these values off their graph, so we assumed an equal weighting on each point using the 0.006 mag error for their reference star measurement. 

In order to take into account the uncertainties in the orbital phase by \citet{shahbaz1993}, we varied the donor fraction in Cen X-4 from 0.80 $<f<$ 1.00 along with a phase-offset that gave the lowest $\chi^2$ for a certain inclination. We obtained an inclination of $(35^{ +4}_{-1})^\circ$ for the binary from which we obtained the mass of the neutron star as $M_{NS} = 1.5^{+.1}_{-.4}M_{\odot}$. Figure 18 shows the best fit ($\chi^{2}_{\nu}$ = 9.57) to the H-band light curve (dashed line) of Cen X-4 for q = 0.20 and i= 36$^\circ$ for a phase-offset of .03, after it was modeled with a K7 V donor plus 6\% constant dilution contribution. The solid line represents the best fit for no phase-offset ($\chi^{2}_{\nu}$ = 12.2). Note, however, that the uncertainty on our final result only reflects the uncertainty in varying the donor star fraction.  It does not include the uncertainty in the light curve measurements.  %Even if we apply the $q = 0.20$ value from Torres et al.\ to the Shahbaz et al.\ light curve models, the latter (their Figure 3) show a 90\% confidence interval of $\sim 34^{\circ}$-- $44^{\circ}$ which is wider than the uncertainty we quote above. 
 As a result, our derived neutron star mass and uncertainty should be viewed with some caution. 

With that caveat, the value $1.1M_{\odot}<M_{NS}<1.6M_{\odot}$ we find is consistent with past estimates found by \citet{casares2007}, \citet{torres2002}, \citet{davanzo05} and \citet{shahbaz1993}. To include the spectral types between M0 V -- K5 V we varied the temperature, T$_{eff}$ from 3700 K to  4500 K, and found that the inclination changes by $\approx 1^\circ$. We obtain the same variation in the inclination if we use $\log{g}$=4.0 if the donor in Cen X-4 were to be a sub-giant. The mass ratio value also has a negligible effect on the inclination. Upon varying the mass ratio within its uncertainties (0.20$\pm$0.03), we found the inclination to change by less than a degree. The biggest uncertainty comes from varying the donor fraction compared to any other parameters.

Both \citet{chevalier89} and \citet{shahbaz1993} have argued that the donor in Cen X-4 is more likely a subgiant than a dwarf. They pointed out that based on the orbital period alone, if the donor were to be a main sequence star, it would underfill the Roche-lobe by a factor of $\approx$ 2.  The CO bands would provide the best tracer of whether the donor star is evolved. We could not test this hypothesis here, both because of the relatively poor data quality in the long wavelength end of the K-band where the strongest CO bandheads occur and because of the dearth of subgiant stars in the IRTF spectral library.  %  Alternatively, the donor can fill its Roche lobe by means of irradiation by hard X-rays from the neutron star during quiescence \citep{hameury1986, podsiadlowski91} in which case we cannot rule out a main sequence star as the donor in Cen X-4. 
Hence, the possibility that the donor in Cen X-4 could be a subgiant (later than K1 IV) remains open at this point . 

%Precise determination of neutron star masses in binary radio-pulsars show a mass distribution centered around 1.35$M_{\odot}$ with an uncertainty of $\pm$.04$M_{\odot}$  \citep{ casares2006}. 
It has been seen that pulsar masses are typically higher ($\approx$ 1.50 M${_\odot}$) in neutron star--white dwarf systems than in neutron star--neutron star systems and this is attributed to higher accretion in these systems \citep{nice2008}. It has been pointed out by \citet{shahbaz1993} that either the neutron star (NS) in Cen X-4 was formed as it reached the end of nuclear evolution in the progenitor's core or it could have collapsed to form a neutron star by gradual accretion on to a dwarf with electron degenerate core. In the latter case, mass of the neutron star could be higher than typical neutron star masses as they can accrete up to several tenths of solar mass during their lifetimes \citep{casares2006,bhatta1991}. Our value of the mass of the neutron star in Cen X-4 exhibits a range $1.1M_{\odot}<M_{NS}<1.6M_{\odot}$ with the most probable value of 1.5$M_{\odot}$. The error on the mass needs to be constrained further in order to realize whether Cen X-4 could be a potential system for understanding nuclear matter and ruling out different equations of state similar to the neutron star LMXB candidate, Cyg X-2 \citep{casares98}.

%\citet{gonzalez2005} have explored the possibility of the donor in Cen X-4 captured matter in the supernova explosion that formed the neutron star. The atmosphere of the donor star should then show anomalous abundances in its composition. In their analysis, they used a $\chi^2$ minimization procedure that provided the best model fit between the spectrum of Cen X-4 and several synthetic spectra and they obtained super-solar abundances for Fe, Ca, Al, {Ti}, {Ni}. They used a K5 V spectral type for the donor in their analysis. We note in particular from our Table 6, that the super-solar abundances of Al and specifically Ca that they derived could be decreased if they moved to a later spectral type star. Our analysis does not rule out a spectral type as late as M1 V for the donor star in Cen X-4. Moving to a cooler star could obviate the requirement of certain super-solar elemental abundances in the donor.

%From our analysis, a super-solar abundance in the donor of Cen X-4 mathematically amounts to obtaining a donor fraction $f>1$. In that case, \ion{Ca}{0} would be super-solar when compared against a  K5 V but not a K7 V or an M0 V spectral type. On the other hand, the donor fraction for \ion{Al}{0} does not require a super-solar abundance when compared to all three spectral types and vice versa for \ion{Fe}{0}. \\

In the analysis done for both V404 Cyg and Cen X-4, there are areas that contribute to systematic uncertainties which may not be accounted for in our final uncertainties. One of these is using a field star to duplicate the Roche-lobe filling donor stars. In V404 Cyg, we found that a K3 III field star with near-solar abundance matched the donor of V404 Cyg quite well except for the \ion{Mg}{1} lines. For Cen X-4, we find that after applying the final donor fraction, some of the lines in the template were still weaker than Cen X-4. Using a field dwarf with solar abundances to represent a Roche-lobe filled star in Cen X-4 could be a plausible explanation for this discrepancy, as could minor mismatches between template and donor star in temperature, gravity, or metallicity. We attempted to mitigate this potential error source in Cen X-4 by averaging over fits to multiple lines and multiple spectral types to obtain the fraction and its uncertainty. Fortunately, our results are dominated by the H-band donor fraction where we obtained consistent results for all three template stars.

Previous studies have assumed that the light curve shape is consistent throughout quiescence.  However, it has been shown in several systems that this assumption is not valid and will lead to systematic errors in the determination of the compact star masses \citep{cantrell2008,cantrell2010}. We obtained the spectra of our targets in 2007 whereas the light curve data for V404 Cyg and Cen X-4 were obtained in 1993 and 1990-91 respectively. This wide time gap between obtaining the light curve data and the spectra adds to the uncertainty associated with the mass of the compact objects.  In V404 Cyg, the accretion disk contribution is very small and has remained so over multiple observations \citep{shahbaz1996,hynes2009}. Our fractional contribution for the donor star agrees with the value obtained by \citet{shahbaz1996} within their statistical uncertainties implying that the small disk contamination in the K-band has changed negligibly in the time gap between the two observations. For Cen X-4, we do not have as much information about the long-term variability of the system. However, the uncertainty on our donor star fraction ($\pm14\%$ in H) and resulting uncertainty on the neutron star mass is broad enough to span the observed range of variability in the system. We also note that the 1.5$M_\odot$ result is consistent with the masses measured in other NS LMXBs. Contemporaneous light curve and spectral data are recommended to obtain a definitive NS mass in Cen X-4.

\section{Conclusion} We have obtained broadband near infrared spectroscopy of V404 Cyg and Cen X-4 in order to directly evaluate the relative contributions in the infrared light due to the donor star and the accretion disk and determine precise compact star masses. We determined the spectral type of the donor star in V404 Cyg to be a K3 III. We performed dilution analysis based on individual prominent metal lines and groups of lines and determined the donor star fraction to be $0.98 \pm .05$ in the H-band and $0.97\pm.09$ in the K-band. With previous light curve data in the H-band, we obtained the inclination of the binary as $i= (67^{+3}_{-1})^\circ$ and a mass for the BH of $9.0^{+.2}_{-.6}$ M$_\odot$ . In Cen X-4 we performed a similar analysis but restricted our fits to the strongest absorption lines due to the lower S/N of the spectrum. We determined the spectral type of the donor star to lie within the range K5 -- M1 V but could not constrain it further. The dilution fraction was determined to be $0.94\pm.14$ in the H-band and  $1.09\pm.20$ in the K-band. From past H-band data we modeled the light curve with a donor star plus a constant flux from the accretion disk and determined the mass of the neutron star in Cen X-4 as $1.5^{+.1}_{-.4} M_{\odot}$ which agrees well with past estimates. These mass measurements should be viewed with some caution however, since contemporaneous light curve and spectral data are required to obtain definitive masses.

%% In this section, we use  the \subsection command to set off
%% a subsection.  \footnote is used to insert a footnote to the text.

%% Observe the use of the LaTeX \label
%% command after the \subsection to give a symbolic KEY to the
%% subsection for cross-referencing in a \ref command.
%% You can use LaTeX's \ref and \label commands to keep track of
%% cross-references to sections, equations, tables, and figures.
%% That way, if you change the order of any elements, LaTeX will
%% automatically renumber them.

%% This section also includes several of the displayed math environments
%% mentioned in the Author Guide.

\acknowledgments{We would like to thank Mattew Beasley for his assistance in observing V404 Cyg and Cen X-4 and  the IRTF staff for their help. We would also like to thank Niall Gaffney for useful discussions.}

Facilities: \facility{IRTF(SpeX)}

\clearpage

\begin{deluxetable}{cccc}
\tabletypesize{\scriptsize}
\tablecaption{Observations}
\tablewidth{0pt}
\tablehead{
\colhead{Object} & \colhead{Date} & \colhead{Exposure time(min)} & \colhead{Orbital Phase}}
\startdata
${\bf  V404 Cyg}$ & 06/07/07 &100 & 0.59-0.61\\
${\bf V404 Cyg}$  & 06/08/07 & 190 & 0.74-0.76 \\
${\bf V404 Cyg}$ & 06/09/07 & 150 & 0.89-0.91 \\
${\bf  Cen X-4}$ & 06/07/07 & 100 & 0.80-0.92\\
${\bf  Cen X-4}$  & 06/08/07 & 30 & -----$^{\tablenotemark{a}}$ \\
${\bf  Cen X-4}$ & 06/09/07 & 110 & 0.97-0.05 \\
\enddata
%% Text for table notes should follow after the \enddata but before
%% the \end{deluxetable}. Make sure there is at least one \tablenotemark
%% in the table for each \tablenotetext.
%%\tablecomments{Table \ref{tbl-1} is published in its entirety in the 
%%electronic edition of the {\it Astrophysical Journal}.  A portion is 
%%shown here for guidance regarding its form and content.}
\tablenotetext{a}{Data from 06/08 for Cen X-4 was not used due to poor data quality at high extinction.}
\end{deluxetable}

%% If you use the table environment, please indicate horizontal rules using
%% \tableline, not \hline.
%% Do not put multiple tabular environments within a single table.
%% The optional \label should appear inside the \caption command.
\clearpage
\begin{deluxetable}{cccc}
\tabletypesize{\scriptsize}
\tablecaption{Equivalent widths for V404 Cyg using FS indices}
\tablewidth{0pt}
\tablehead{
\colhead{Feature} & \colhead{Symbol} & \colhead{Integration limits} & \colhead{Equivalent width($\AA$)}}
\startdata
${\bf  SiI }$ ${1.5892}$ & ${\bf W_{1.59}}$ &1.5870-1.5910 & 3.36 $\pm$ 0.07\\
${\bf  ^{12}CO(6,3) }$ ${1.6187}$ & ${\bf W_{1.62}}$ &1.6175-1.6220 & 4.10 $\pm$ 0.08\\
${\bf  Na I }$ ${2.2076}$ & ${\bf W_{Na}}$ &2.2053-2.2101 & 2.03 $\pm$ 0.10\\
${\bf  FeI }$ ${2.2263}$ & ${\bf W_{Fe1}}$ &2.2248-2.2293 & 1.22 $\pm$ 0.74\\
${\bf  FeI }$ ${2.2387}$ & ${\bf W_{Fe2}}$ &2.2367-2.2402 & 0.76 $\pm$ 0.12\\
${\bf  CaI}$ ${2.2636}$ & ${\bf W_{Ca}}$ &2.2611-2.2662 & 2.06 $\pm$ 0.23\\
${\bf  MgI }$ ${2.2814}$ & ${\bf W_{Mg}}$ &2.2788-2.2840 & 1.18 $\pm$ 0.18\\
${\bf  ^{12}CO(2,0) }$ ${2.2935}$ & ${\bf W_{2.29}}$ &2.2924-2.2977 & 9.11 $\pm$ 0.76\\
${\bf  ^{12}CO(3,1)}$ ${2.3227}$ & ${\bf W_{2.32}}$ &2.3218-2.3272 & 9.15 $\pm$ 0.28\\
${\bf  ^{13}CO(2,0) }$ ${2.3448}$ & ${\bf W_{2.34}}$ &2.3436-2.3491 & 3.30 $\pm$ 0.56\\
\enddata
\tablecomments{The equivalent widths were calculated using the integration limits from \citet{fs2000}. The error bar on the measurements are based on continuum placement uncertainties.}
\end{deluxetable}

%% If the table is more than one page long, the width of the table can vary
%% from page to page when the default \tablewidth is used, as below.  The
%% individual table widths for each page will be written to the log file; a
%% maximum tablewidth for the table can be computed from these values.
%% The \tablewidth argument can then be reset and the file reprocessed, so
%% that the table is of uniform width throughout. Try getting the widths
%% from the log file and changing the \tablewidth parameter to see how
%% adjusting this value affects table formatting.

%% The \dataset{} macro has also been applied to a few of the objects to
%% show how many observations can be tagged in a table.

%% You can append references to a table using the \tablerefs command.
%%\tablerefs{
%%(1) Barbuy, Spite, \& Spite 1985; (2) Bond 1980; (3) Carbon et al. 1987;
%%(4) Hobbs \& Duncan 1987; (5) Gilroy et al. 1988: (6) Gratton \& Ortolani 1986}

\clearpage
\begin{deluxetable}{ccccccccc}
\tabletypesize{\scriptsize}
\tablecaption{Donor fractions, $f$, obtained for V404 Cyg with K2 III - K4 III field stars}
\tablewidth{0pt}
\tablehead{
\colhead{Band} & \colhead{Wavelength range($\mu$m)} & \colhead{Feature} & \colhead{$f$(K2 III) } & \colhead{ $\chi^{2}_{\nu}$} &  \colhead{$f$(K3 III)} & \colhead{$\chi^{2}_{\nu}$} & \colhead{$f$(K4 III)} & \colhead{$\chi^{2}_{\nu}$}}
\startdata
&${\bf  1.97-2.42 }$ &entire K-band&1.05 & 8.90 &0.95 & 5.83 &0.84 & 8.79 \\
&${\bf 1.97-1.99 }$& Ca I  &1.36&2.55& 1.07&4.50& 1.18&9.53\\
&${\bf  2.112-2.128 }$ &Al I &1.50 &  0.20 &1.05 & 0.52 &1.57 &0 .35 \\
%${\bf K}$&${\bf  2.14-2.15 }$& Mg I &  1.04&0.14& 0.99&0.42& 1.92&0.17\\
${\bf K}$&${\bf  2.203-2.214 }$&Na I  & 1.50&0.24& 0.94&0.15& 1.07&0.23 \\
%&${\bf  2.220-2.245 }$ &Fe I & 1.06&1.69& 0.95&0.77& 0.90&1.00\\
&${\bf  2.26-2.27 }$ &Ca I&1.24&0.21&0.85&0.15&0.98&0.26\\
&${\bf  2.276-2.286 }$ & Mg I&1.60 &0.89 &1.25 &0.87&1.87 &0.73\\
&${\bf  2.28-2.42 }$& CO bands &1.05&13.0&0.95&6.70&0.87&17.80\\
& &&&&&&&\\
\hline
&&&&&&&&\\
&${\bf 1.48-1.72 }$& entire H-band & 1.33&21.2& 1.09&13.42& 1.05&18.76\\
&${\bf  1.48-1.52 }$& Mg I  & 1.53&3.57& 1.19&4.34& 1.40&5.96\\
&${\bf 1.574-1.587}$& Mg I &1.43&7.00&1.15&4.99&1.19&6.78\\
${\bf H}$&${\bf 1.584-1.595 }$&Si I & 1.07&3.70& 0.94&2.44& 0.98&2.50\\
&${\bf 1.615-1.650 }$& CO bands  &1.24&10.5&0.98&6.33&0.95&9.00 \\
&${\bf 1.670-1.678 }$& Al I & 1.32&3.77& 1.03&2.14& 1.17&1.85\\
&${\bf  1.70-1.72 }$& Mg I  & 1.56&4.00& 1.18&5.57& 0.98&5.00\\
&&&&&&&&\\
\hline
&&&&&&&&\\
%&${\bf 0.8-1.1 }$&part of J-band &0.92 &1.80&0.70 &0.25& 0.86& 2.18 \\
&${\bf 0.854-0.857}$               &Mg I              & 1.02              &1.27           &0.91           &0.51             &0.95              &1.05\\
&${\bf 0.865-0.869}$               &Ca I              & 1.03             &1.71              &0.89           &1.62              &0.93              &2.15\\
&${\bf 1.158-1.167 }$              & Fe I              &1.11              &0.08              &0.83           &0.10             & 1.09             &0.12\\
&${\bf 1.168-1.172 }$              & K I                &1.23              &0.04              &0.96           &0.07             & 1.11             &0.08\\
&${\bf 1.182-1.186 }$              & Mg I            &1.33               &0.19              &1.12           &0.09             & 1.24             &0.05\\
&${\bf 1.187-1.192 }$              & Fe I             &1.08                &0.09              &0.85          &0.02             & 0.90             &0.03\\
${\bf J}$&${\bf 1.202-1.207 }$& Si I              &1.23             &0.04                &1.05           &0.02            & 1.17              &0.03\\
%&${\bf 1.208-1.210 }$& Mg I&1.05 &0.04&1.13 &0.02& 1.37&0.02\\
&${\bf 1.279-1.287 }$              & Ti I               &1.59               &0.23              &1.38            &0.21            & 1.59              &0.40\\
&${\bf 1.287-1.292 }$              & Mn I             &0.99               &0.05             &0.88            &0.02            & 0.95              &0.03\\
&${\bf 1.310-1.317 }$              & Al I              &1.23                &0.15              &1.09            &0.10             & 1.16            &0.08\\
&&&&&&&&
\enddata
%\tablecomments{The donor fraction, $f$, calculated with K2 III - K4 III templates for V404 Cyg is listed in table 4 along with the rms values obtained. }
\end{deluxetable}

\clearpage
\begin{deluxetable}{cccc}
\tabletypesize{\scriptsize}
\tablecaption{Equivalent Width analysis for Cen X-4 using FS indices}
\tablewidth{0pt}
\tablehead{
\colhead{Feature} & \colhead{Symbol} & \colhead{Integration limits} & \colhead{Equivalent width($\AA$)}}
\startdata
${\bf  SiI }$ ${1.5892}$ & ${\bf W_{1.59}}$ &1.5870-1.5910 & 3.08 $\pm$ 0.41\\
${\bf  ^{12}CO(6,3) }$ ${1.6187}$ & ${\bf W_{1.62}}$ &1.6175-1.6220 & 1.90 $\pm$ 0.13\\
${\bf  Na I }$ ${2.2076}$ & ${\bf W_{Na}}$ &2.2053-2.2101 & 2.56 $\pm$ 0.21\\
${\bf  FeI }$ ${2.2263}$ & ${\bf W_{Fe1}}$ &2.2248-2.2293 & 2.25 $\pm$ 0.36\\
${\bf  FeI }$ ${2.2387}$ & ${\bf W_{Fe2}}$ &2.2367-2.2402 & 0.51 $\pm$ 0.38\\
${\bf  CaI}$ ${2.2636}$ & ${\bf W_{Ca}}$ &2.2611-2.2662 & 3.30 $\pm$ 0.55\\
${\bf  MgI }$ ${2.2814}$ & ${\bf W_{Mg}}$ &2.2788-2.2840 & 0.89 $\pm$ 0.48\\
${\bf  ^{12}CO(2,0) }$ ${2.2935}$ & ${\bf W_{2.29}}$ &2.2924-2.2977 & 4.50 $\pm$ 0.44\\
${\bf  ^{12}CO(3,1)}$ ${2.3227}$ & ${\bf W_{2.32}}$ &2.3218-2.3272 & 3.01 $\pm$ 0.43\\
${\bf  ^{13}CO(2,0) }$ ${2.3448}$ & ${\bf W_{2.34}}$ &2.3436-2.3491 & 1.0 $\pm$ 0.59\\
\enddata
\tablecomments{The equivalent widths were calculated using the integration limits from \citet{fs2000}. The error bar on the measurements are based on continuum placement uncertainties.}
\end{deluxetable}

\clearpage
\begin{deluxetable}{cccc}
\tabletypesize{\scriptsize}
\tablecaption{Equivalent Width analysis for Cen X-4 using Ali et al.\ indices.}
\tablewidth{0pt}
\tablehead{
\colhead{Feature} & \colhead{Symbol} & \colhead{Integration limits} & \colhead{Equivalent width($\AA$)}}
\startdata
${\bf  Na I }$ ${2.2076}$ & ${\bf W_{Na}}$ &2.204-2.211 & 3.33 $\pm$ 0.33\\
${\bf  CaI}$ ${2.2636}$ & ${\bf W_{Ca}}$ &2.258-2.269 & 5.89 $\pm$ 1.23\\
${\bf  MgI }$ ${2.2814}$ & ${\bf W_{Mg}}$ &2.279-2.285 & 1.65 $\pm$ 0.57\\
${\bf  ^{12}CO(2,0) }$ ${2.2935}$ & ${\bf W_{2.29}}$ &2.289-2.302 & 5.27 $\pm$ 1.09\\
\enddata
\tablecomments{The equivalent widths were calculated using the integration limits from Ali,1995. The error bar on the measurements are based on continuum placement uncertainties.}
\end{deluxetable}

\clearpage
\begin{deluxetable}{ccccccccc}
\tabletypesize{\scriptsize}
\tablecaption{Donor fractions , $f$,  calculated for Cen X-4 with K5 V - M0 V field stars}
\tablewidth{0pt}
\tablehead{
\colhead{Band}&\colhead{Feature} & \colhead{Wavelength range($\mu$m)} &  \colhead{$f$ (K5 V)} & \colhead{ $\chi^{2}_{\nu}$} &   \colhead{$f$ (K7 V)} &  \colhead{$\chi^{2}_{\nu}$}  &  \colhead{$f$ (M0 V)}   &  \colhead{$\chi^{2}_{\nu}$} }
\startdata
&                          ${\bf 1.97-2.42 }  $      &K-band             &1.00           &6.76         & 0.90        &7.40       & 0.93      &6.76   \\
&                          ${\bf 1.97-1.99 }  $      &Ca I              &1.20           &2.62         & 0.85        &3.72       & 0.73      &3.83   \\
%&                          ${\bf 2.108-2.125 }  $     &Al I            &0.85           &1.18         & 0.93        &1.05       &0.85       &1.02 \\
${\bf K}$&                          ${\bf  2.203-2.214 }  $     &Na I            & 1.48          &1.39         & 1.25        &1.05       &1.15       &1.15 \\
&                         ${\bf  2.260-2.272 }  $        &Ca I       &1.15           &0.95         & 0.98        &1.16        &0.95      &1.17  \\
%&Mg I&                    ${\bf  2.10-2.125 }    $                & 0.88           &.015        & 0.83        &.018        &0.78      &.015  \\
&           ${\bf  2.295-2.305}   $       &$^{12}$CO         &1.20            &1.38        &  1.12       &1.60        &  1.00    &1.67  \\
&&&&&&&&\\
\hline
&&&&&&&&\\
&                          ${\bf 1.48-1.72 }  $      &H-band              &0.89           &2.56         & 0.93        &2.75       & 0.92      &2.76   \\
&                         ${\bf 1.48-1.52 }$    &Mg I                  & 0.93           &1.24        &  0.97      &1.07         & 0.98     &1.20   \\
&			${\bf 1.57-1.582}$ 	&Mg I	      &0.95		&0.90         & 1.04	 &1.02	  & 0.98	&0.85  \\
${\bf H}$&			${\bf 1.585-1.605}$ 	&Si I	      &1.10		&1.37 	  & 1.15	 &1.60	  & 1.22	&1.33  \\
&${\bf  1.615-1.65 }$ 	&$^{12}$CO	      & 0.93		&1.02	 & 0.93 	&0.96	& 0.92	&1.20 \\
&			${\bf 1.67-1.68}$ 	&Al I	     &0.88 		&2.46	&  0.93	&2.59		& 0.90       &2.54\\
&			${\bf  1.70-1.72 }$ 	&Mg I	     & 0.71 	 	&2.25	& 0.71	&2.21	& 0.68	&2.12  \\
&&&&&&&&\\
\hline
&&&&&&&&\\
&${\bf 1.182-1.186 }$              & Mg I            &0.92               &0.70             &1.03           &0.56            & 1.06             &0.72\\
&${\bf 1.187-1.192 }$              & Fe I             &1.26                &1.01              &1.15          &1.02             & 1.23             &1.06\\
${\bf J}$&${\bf 1.202-1.207 }$& Si I              &1.38             &0.88                &1.71           &1.06            & 1.82              &1.28\\
&${\bf 1.310-1.317 }$              & Al I              &0.86               &1.55              &0.86           &1.92            & 0.80           &0.93\\
&&&&&&&&\\
\enddata
%\tablecomments{The donor fractions calculated with K5 V - M0V templates for Cen X-4 is given with the rms values obtained.}
\end{deluxetable}

\clearpage

\begin{figure}
\figurenum{1}
\epsscale{1.0}
\plotone{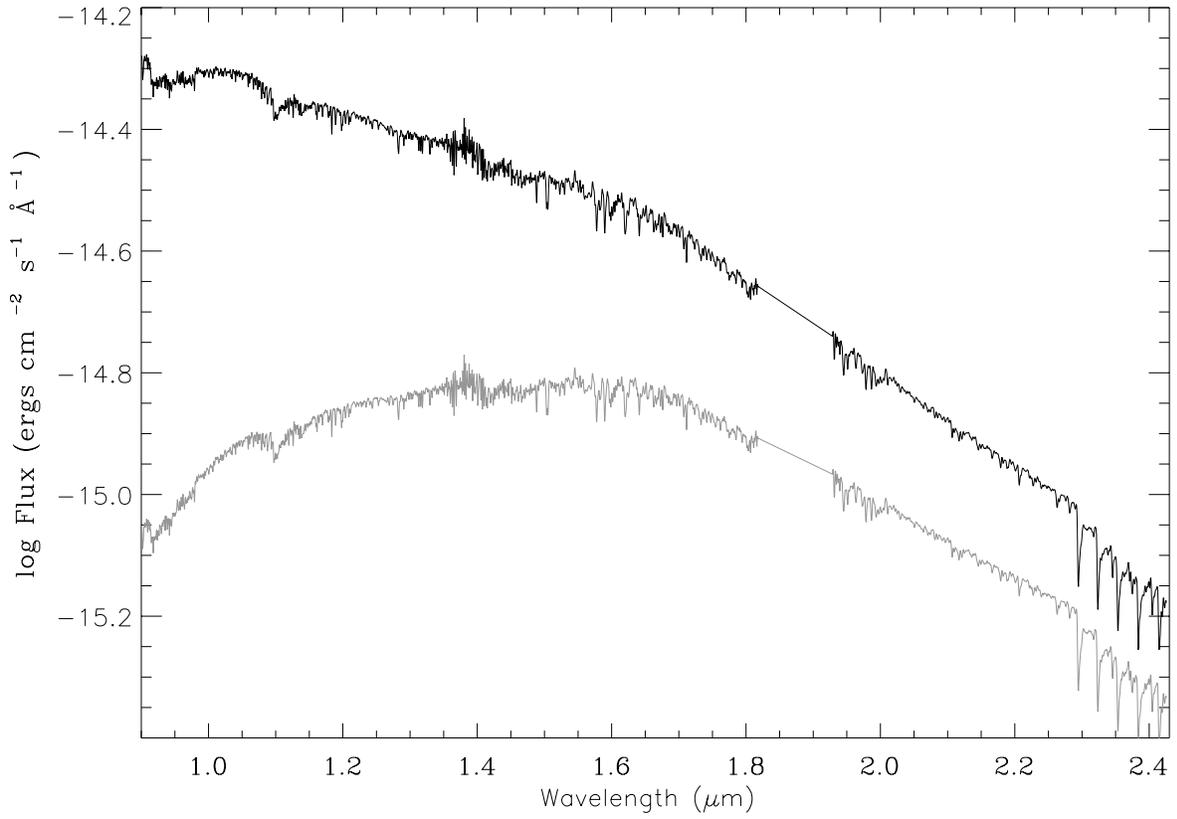}
\caption{The NIR spectrum of V404 Cyg. The time-averaged spectrum, shown in gray, was obtained after correcting for atmospheric absorption and shifting the individual exposures to the rest frame of the donor star. The dereddened spectrum, using E(B-V)=1.303 \citep{hynes2009}, is shown in black.}
\end{figure}

\clearpage

\begin{figure}
\figurenum{2}
\epsscale{1.0}
\plotone{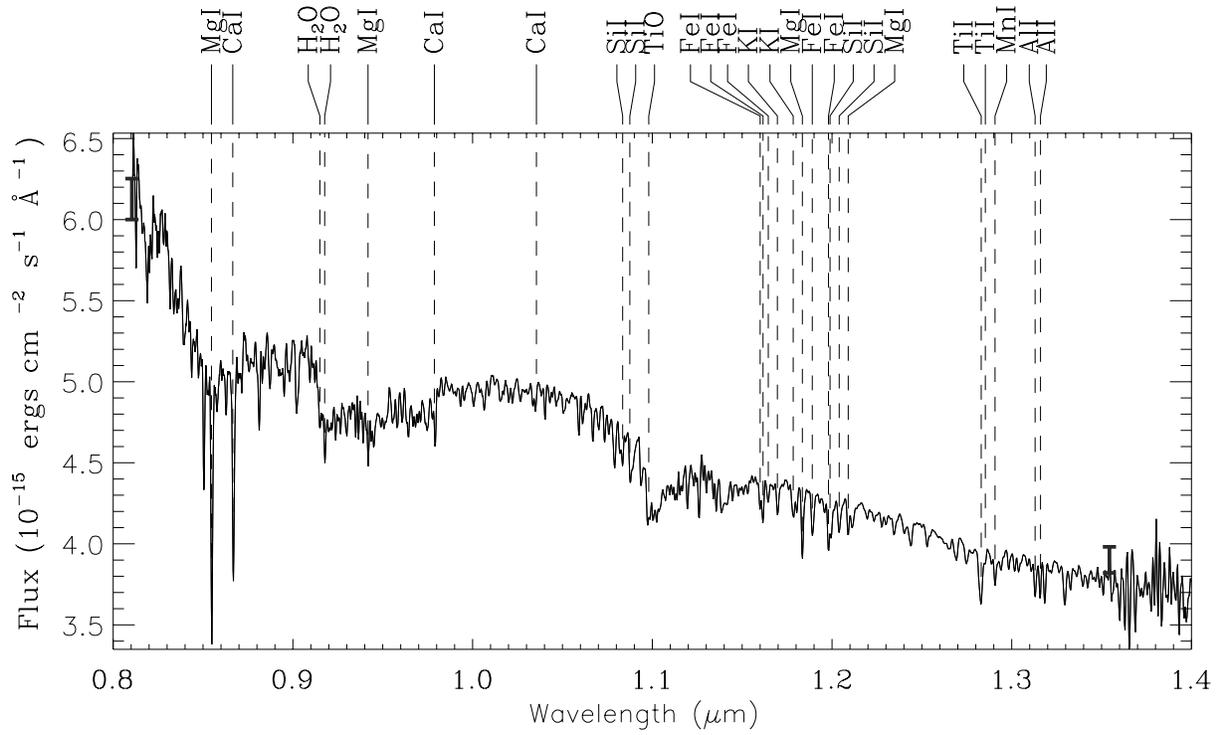}
\caption{The J-band spectrum of V404 Cyg. Prominent spectral features are labeled at the top of the figure and representative error bars are shown near .81$\mu$m and 1.35 $\mu$m.}
\end{figure}

\clearpage

\begin{figure}
\figurenum{3}
\epsscale{1.0}
\plotone{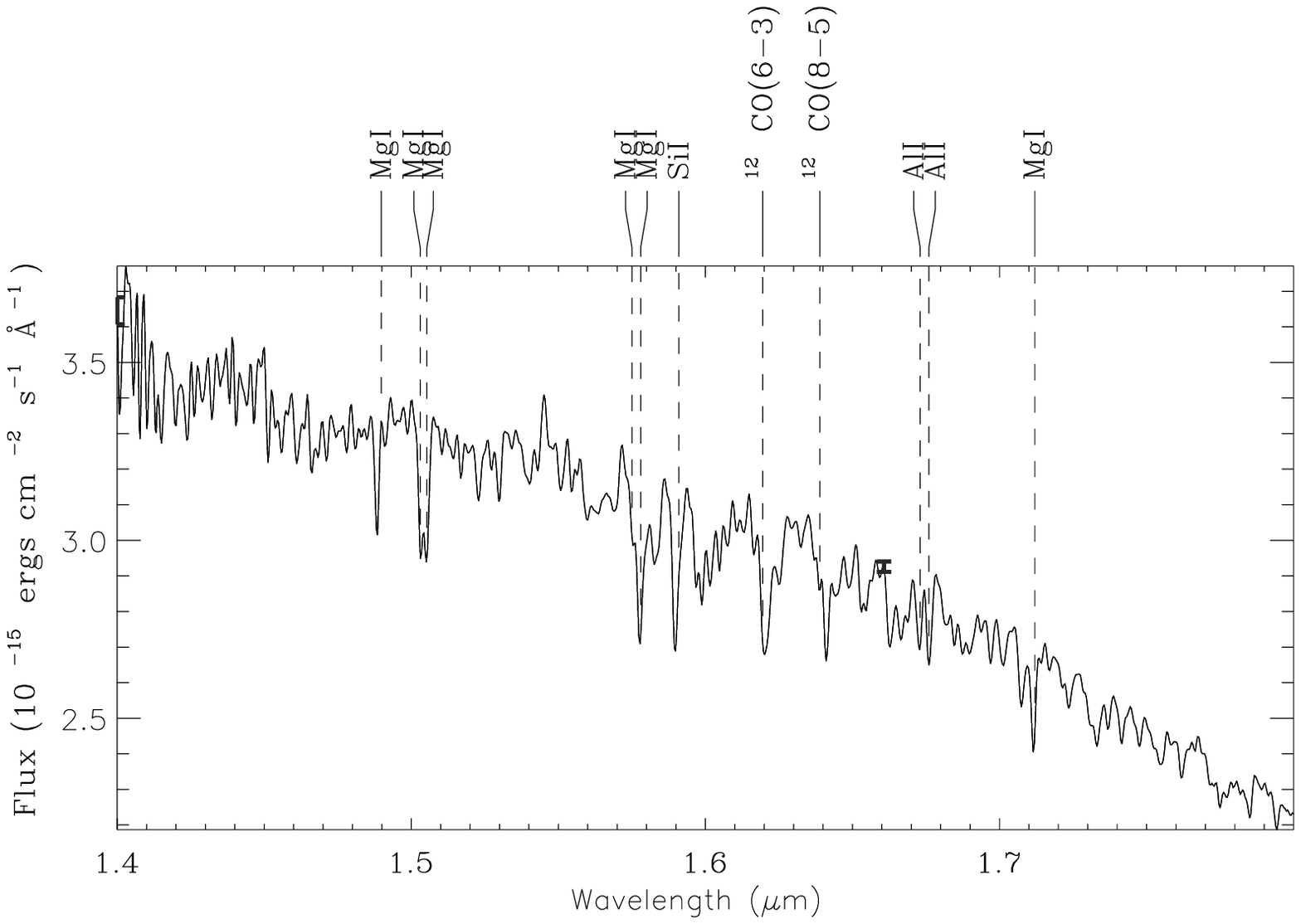}
\caption{The H-band spectrum of V404 Cyg with representative error bars.}
\end{figure}

\clearpage

\begin{figure}
\figurenum{4}
\epsscale{1.0}
\plotone{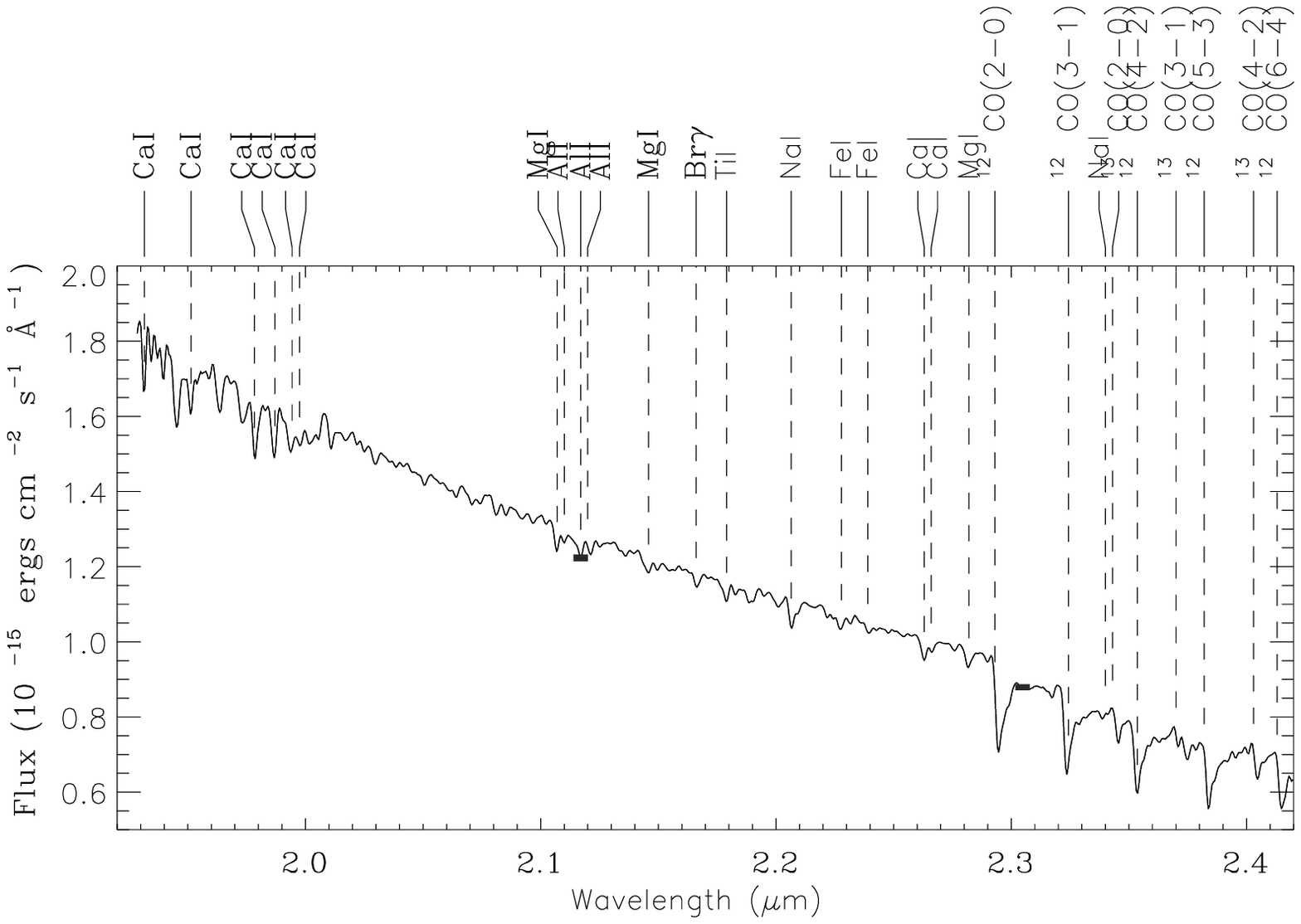}
\caption{The K-band spectrum of V404 Cyg with representative error bars. }
\end{figure}

\clearpage

\begin{figure}
\figurenum{5}
\epsscale{1.0}
\plotone{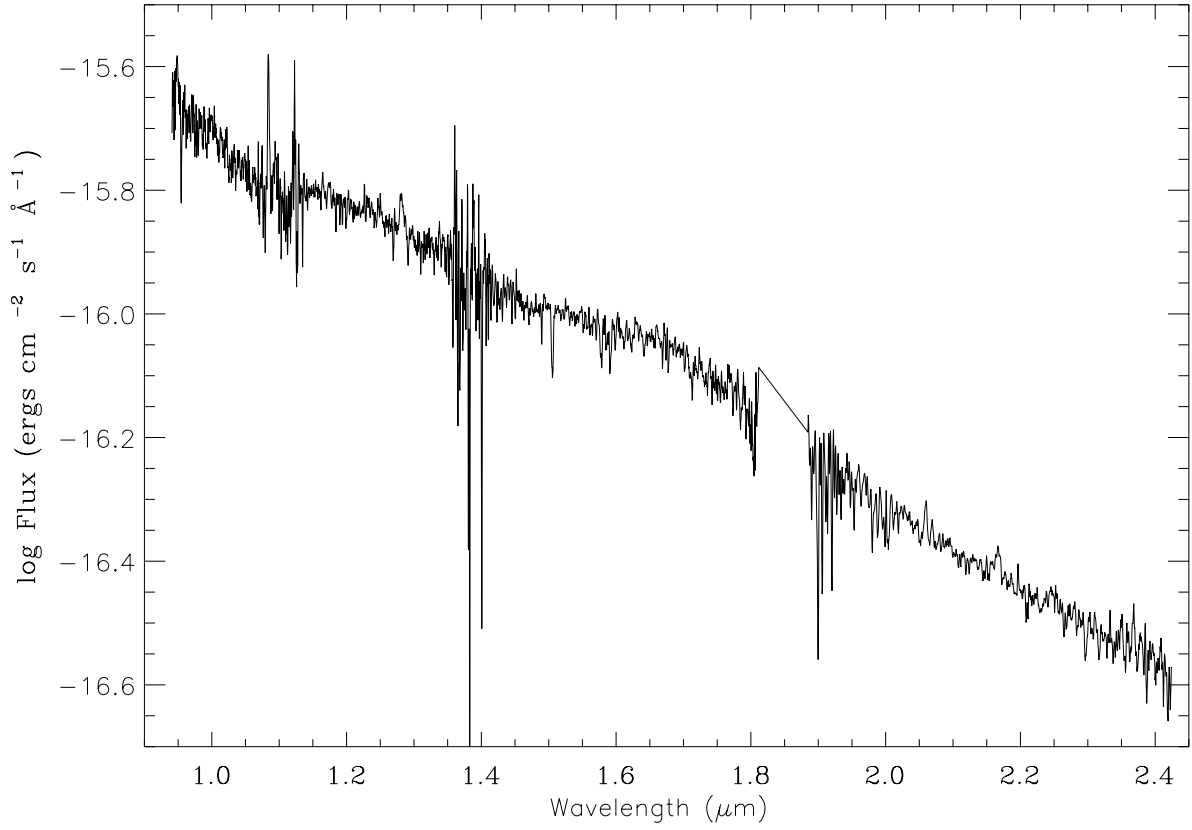}
\caption{The NIR spectrum spectrum of Cen X-4. The time averaged spectrum was obtained by correcting the individual exposures for atmospheric absorption and shifting them to the rest frame of the donor star.  The spectrum has been dereddened using E(B-V)=0.1 \citep{blair84}.}
\end{figure}

\clearpage

\begin{figure}
\figurenum{6}
\epsscale{1.0}
\plotone{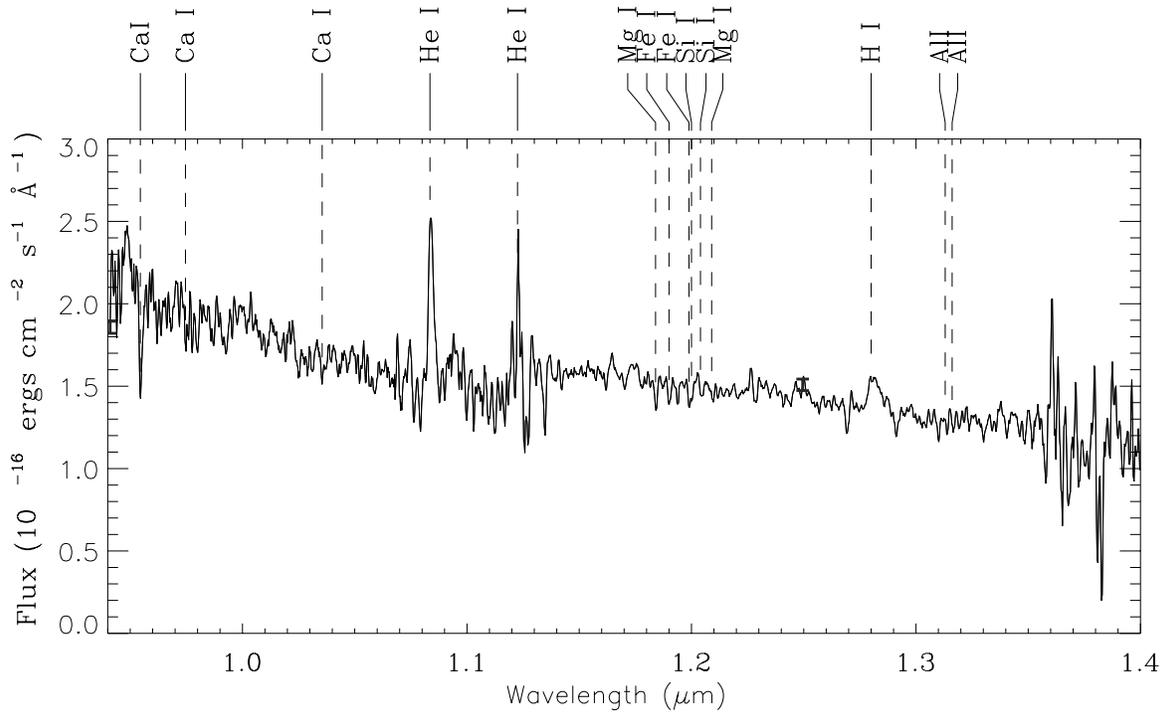}
\caption{The J-band spectrum of Cen X-4. Prominent emission and absorption features are labeled at the top of the figure. Representative error bars are shown near wavelengths .94 $\mu$m and 1.25 $\mu$m.}
\end{figure}

\clearpage

\begin{figure}
\figurenum{7}
\epsscale{1.0}
\plotone{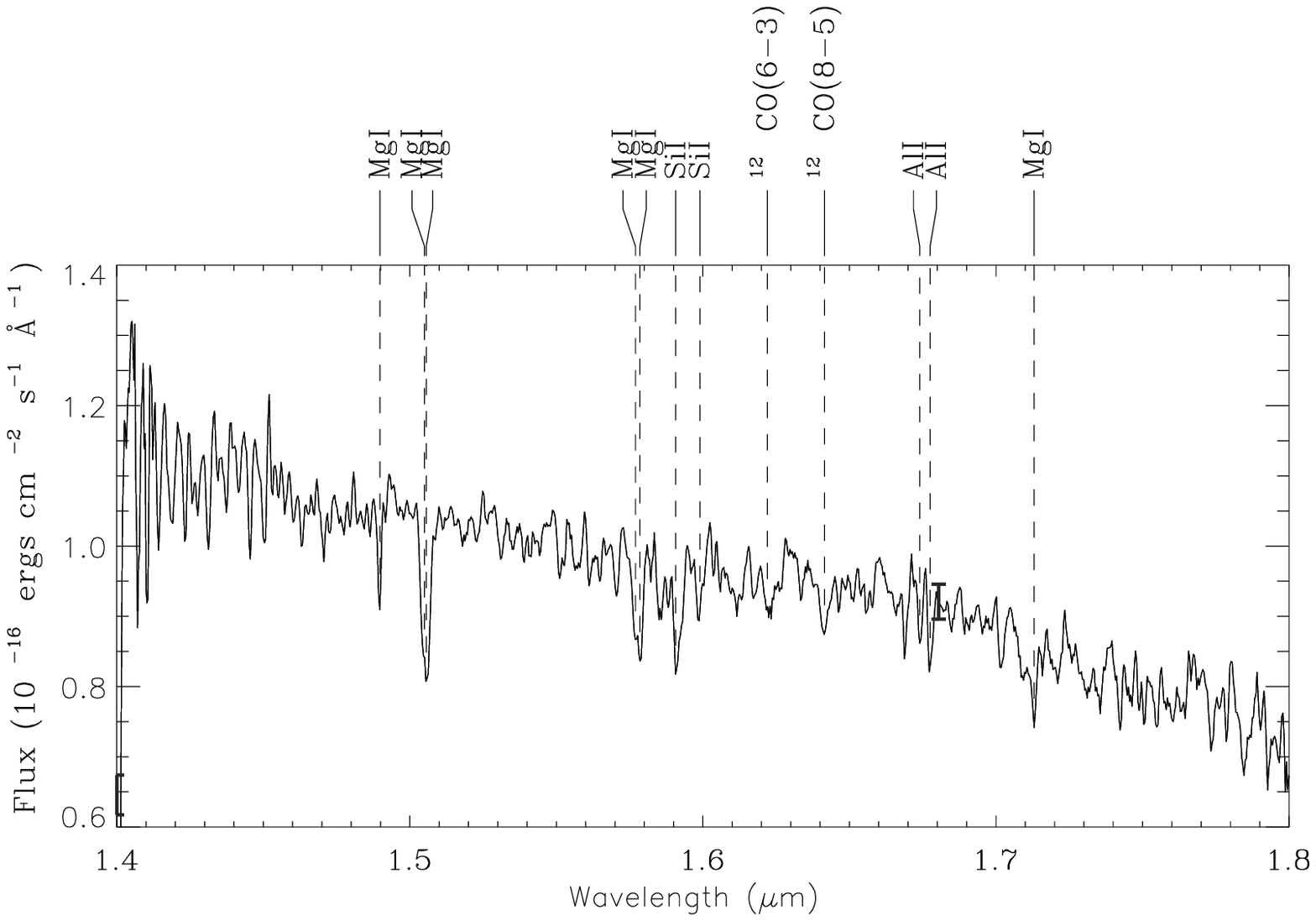}
\caption{The H-band spectrum of Cen X-4 with representative error bars.}
\end{figure}

\clearpage

\begin{figure}
\figurenum{8}
\epsscale{1.0}
\plotone{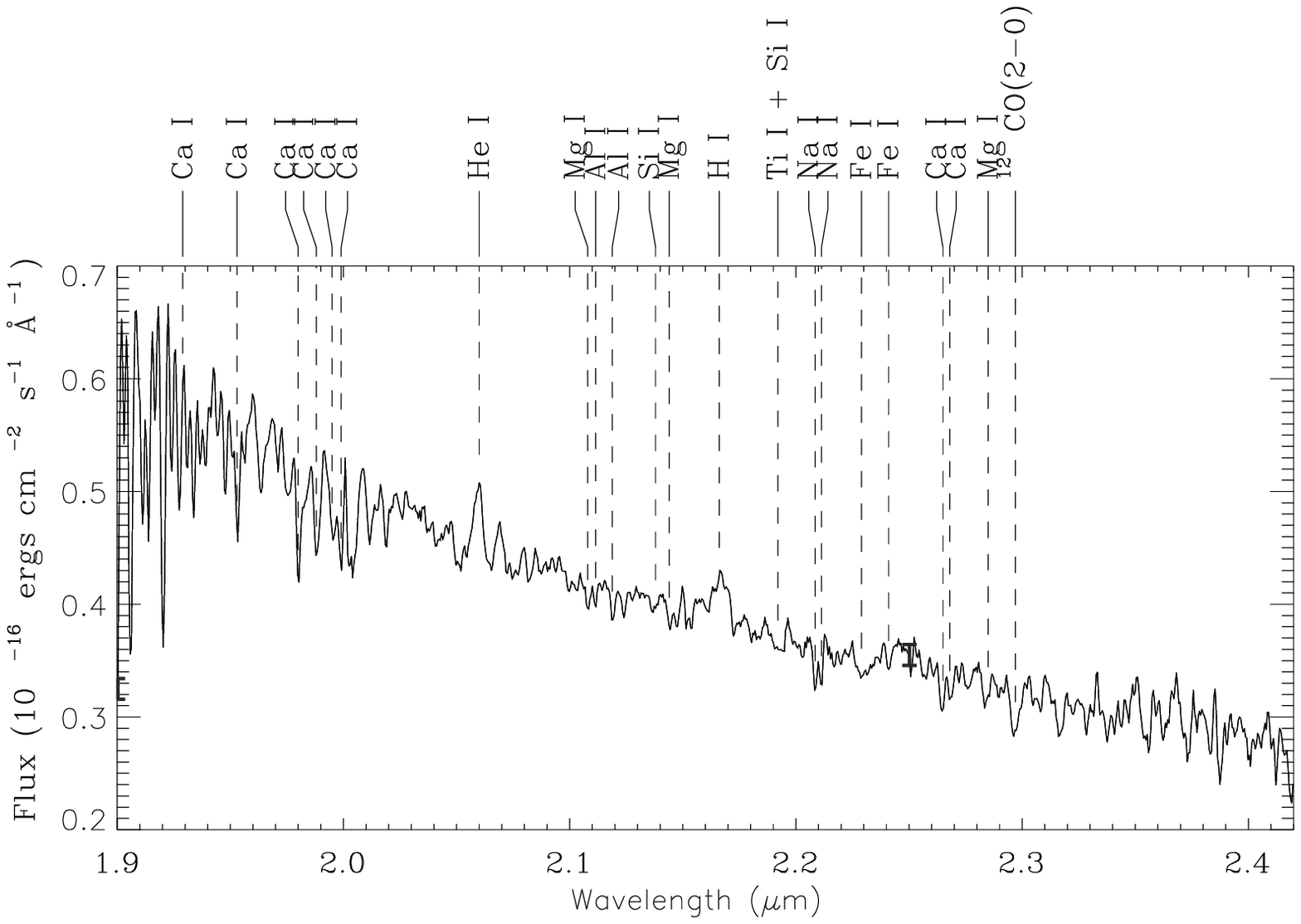}
\caption{The K-band spectrum of Cen X-4 with representative error bars.}
\end{figure}

\clearpage

\begin{figure}
\figurenum{9}
\epsscale{1.0}
\plotone{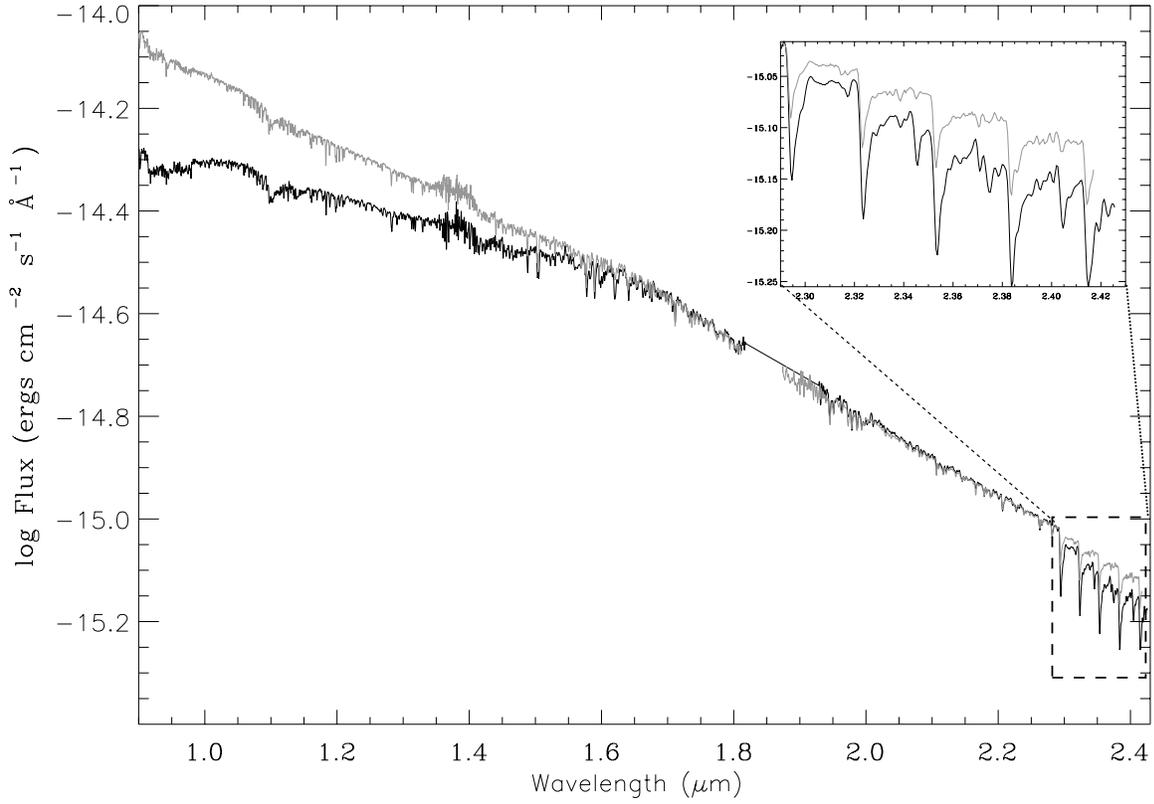}
\caption{Comparison of V404 Cyg ( shown in black) with a K1 IV field star (shown in gray). The template has been normalized to the spectrum of V404 Cyg just blueward of the 2.29$\mu$m $^{12}$CO absorption bandhead. A zoomed in version of the CO bands (2.29 -- 2.42 $\mu$m) shows that a K1 IV star does not match the relative depths of the CO bands in V404 Cyg.}
\end{figure}

\clearpage

\begin{figure}
\figurenum{10}
\epsscale{1.0}
\plotone{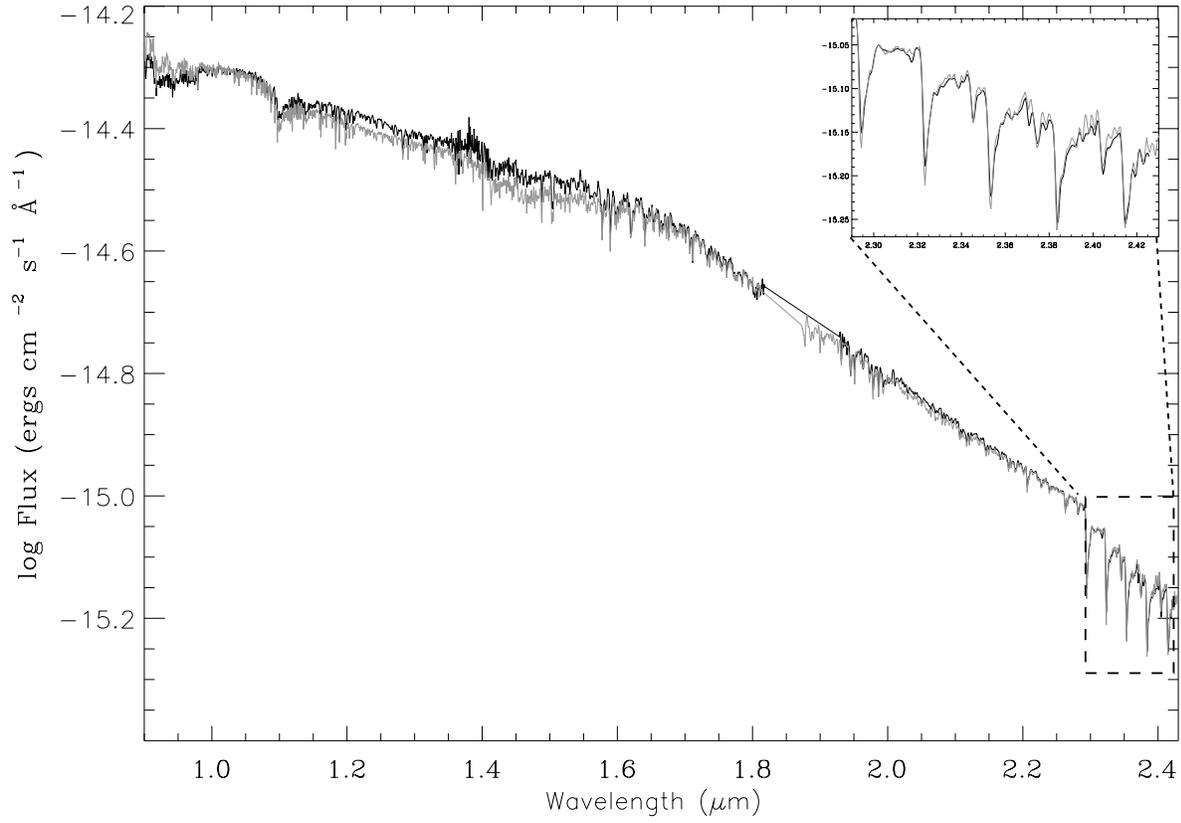}
\caption{Comparison of V404 Cyg (shown in black) with a K3 III field star (shown in gray). The zoomed in version of the CO bands shows a better match between a K3 III star and V404 Cyg with respect to the depths of these absorption features.}
\end{figure}

\clearpage

\begin{figure}
\figurenum{11}
\epsscale{1.0}
\plotone{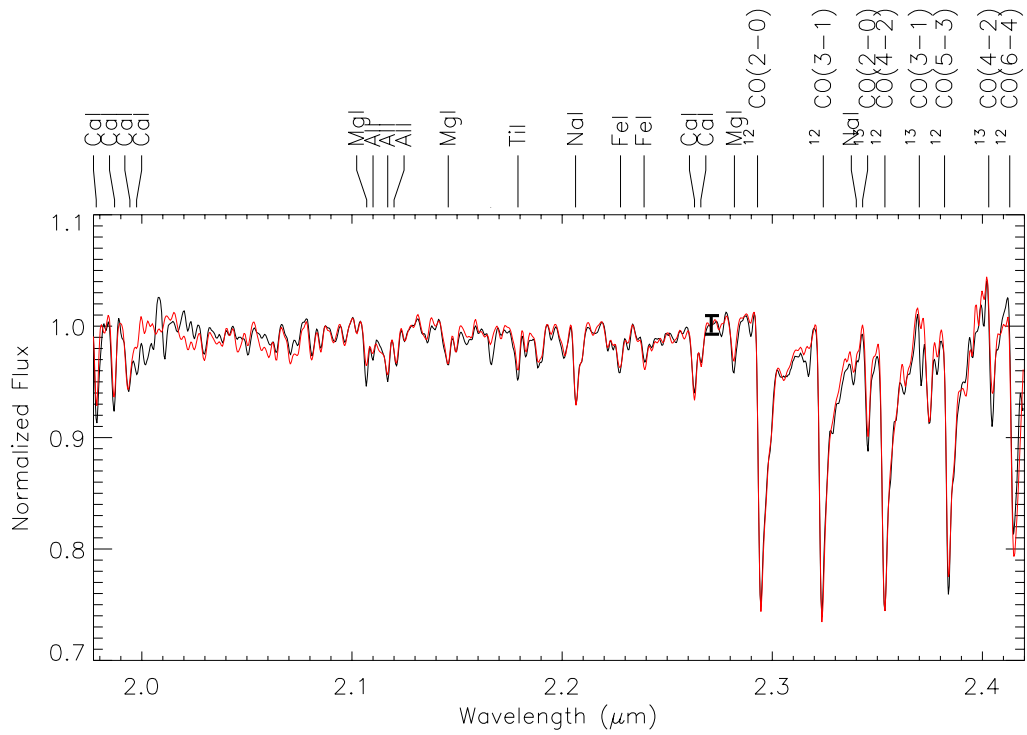}
\caption{The normalized Kband spectrum of V404 Cyg is shown in black. Overplotted in red is a K3 III field star normalized and scaled to $f=.97$ of K3 III field star. Representative error bars are shown in the figure.}
\end{figure}

\clearpage

\begin{figure}
\figurenum{12}
\epsscale{1.0}
\plotone{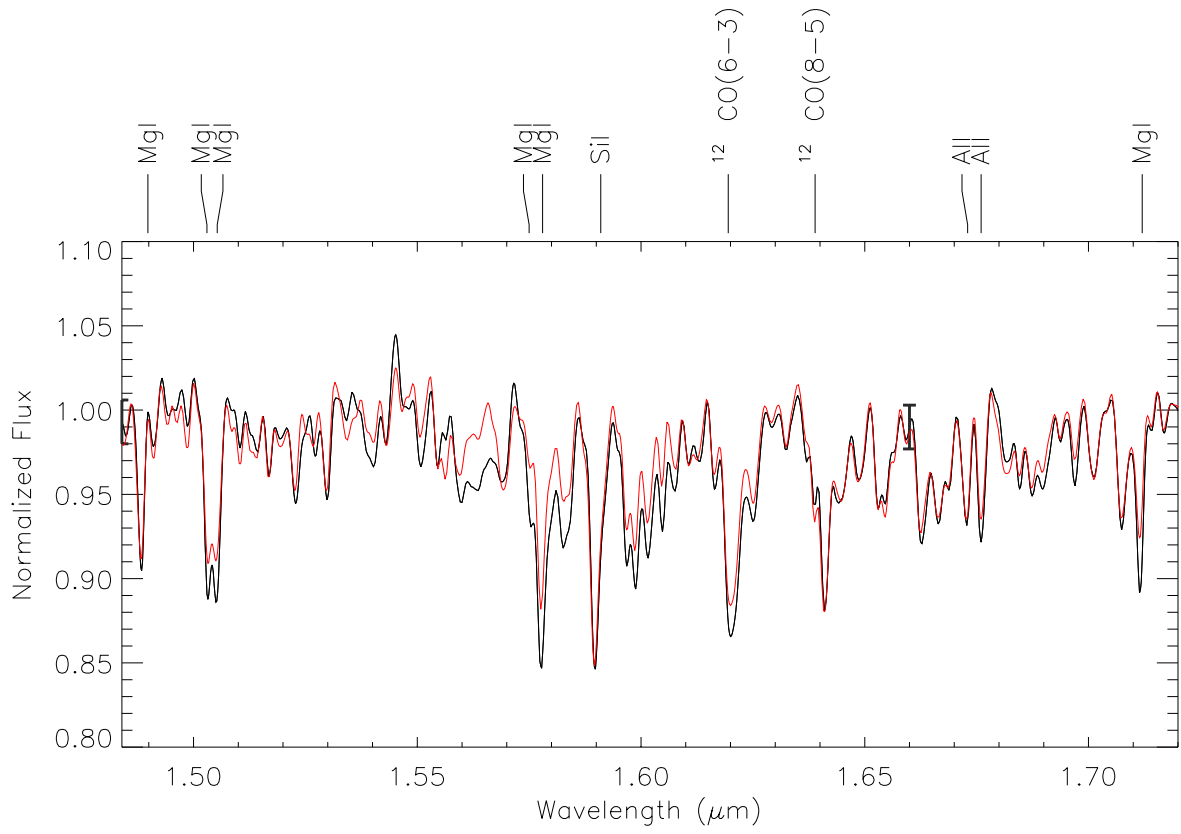}
\caption{The normalized H-band spectrum of V404 Cyg is shown in black. Overplotted in red is a K3 III field star normalized and scaled to $f=.98$. Representative error bars are shown in the figure.}
\end{figure}

\clearpage

\begin{figure}
\figurenum{13}
\epsscale{1.0}
\plotone{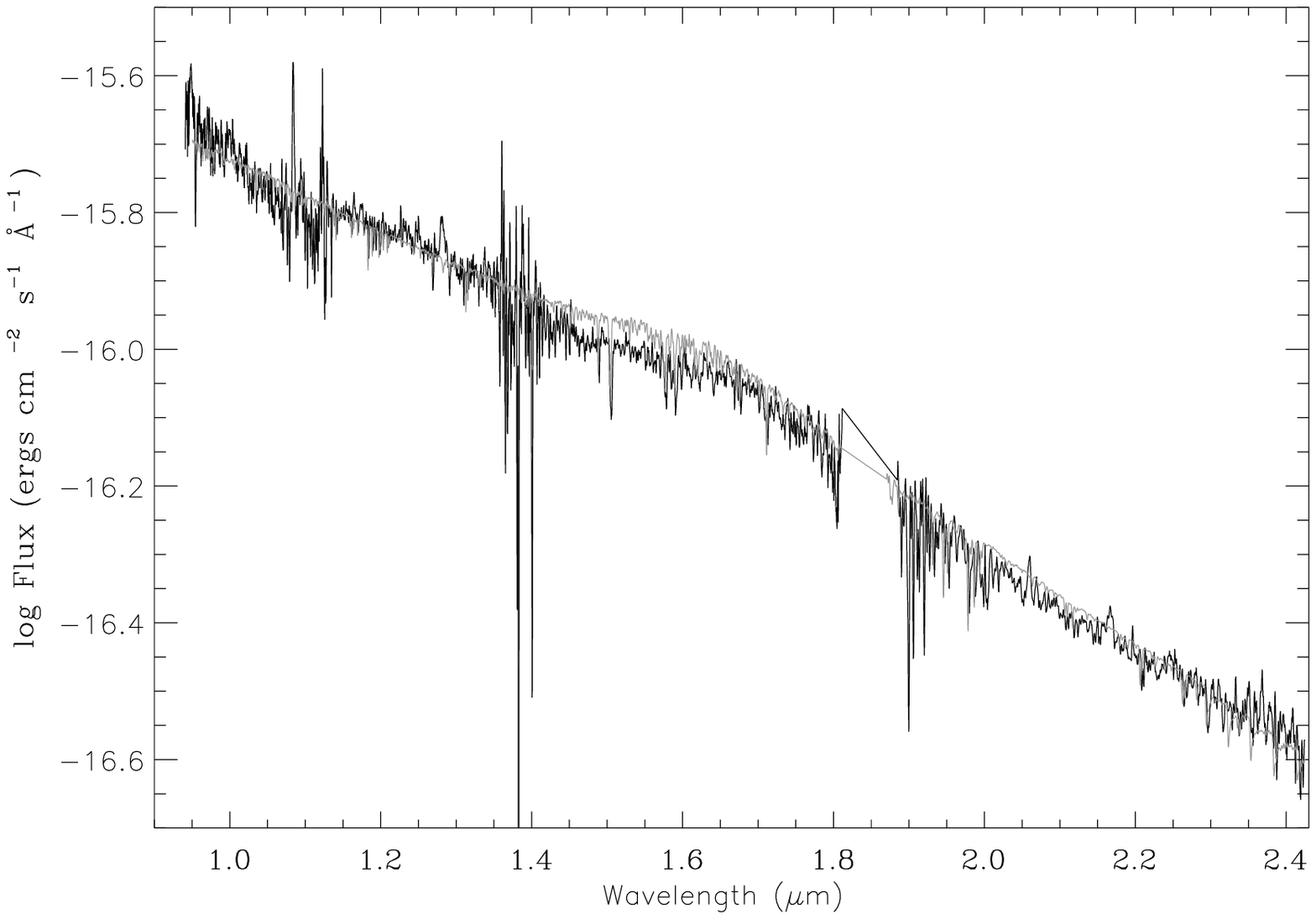}
\caption{Comparison of the NIR spectrum of Cen X-4 (shown in black) with a K7 V field star  (shown in gray).  The field star spectrum has been normalized to that of Cen X-4 just blueward of  $^{12}$CO bandhead at 2.29 $\mu$m. }
\end{figure}

\clearpage

\begin{figure}
\figurenum{14}
\epsscale{1.0}
\plotone{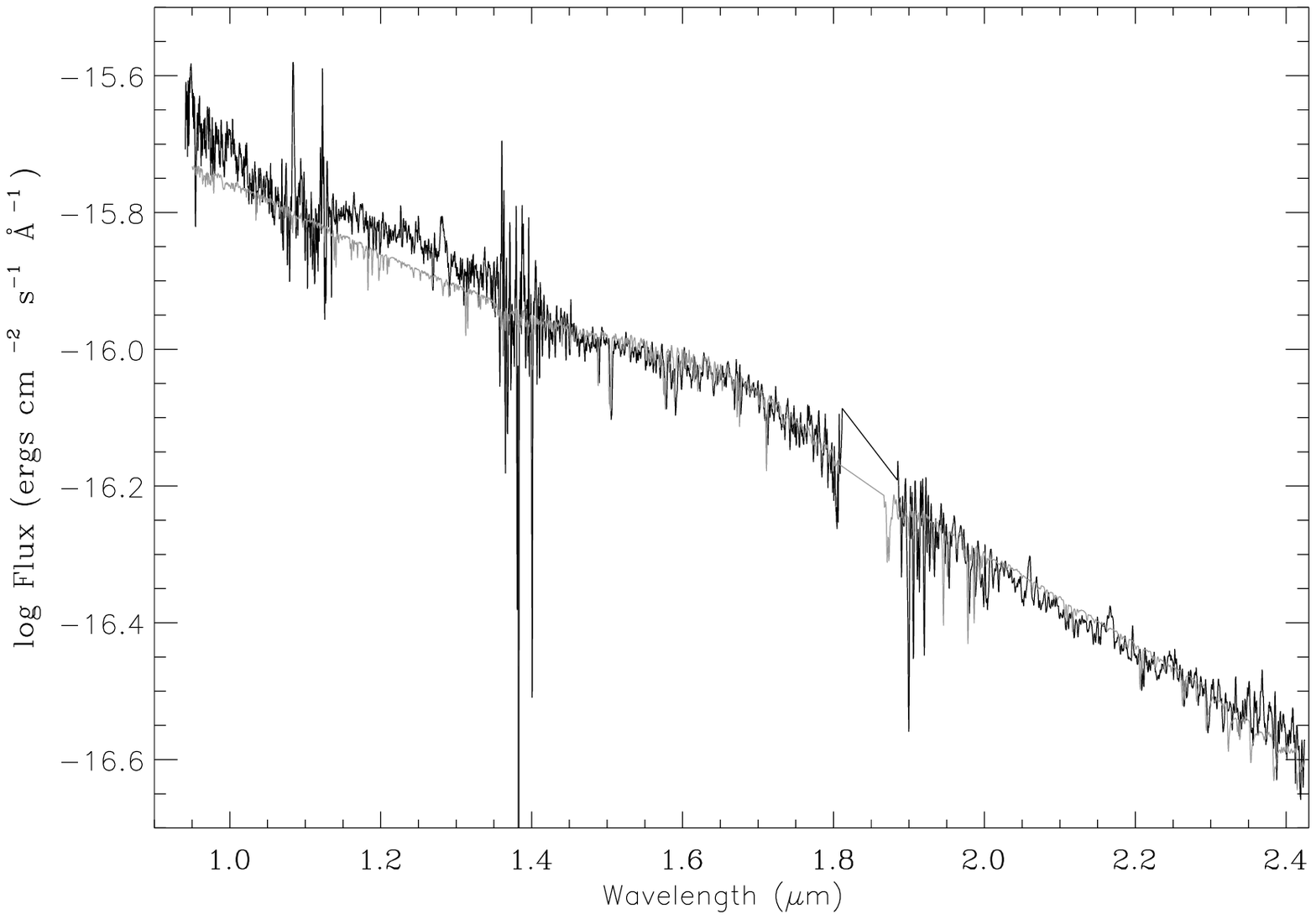}
\caption{Comparison of the NIR spectrum of Cen X-4 (shown in black) with a M0V field star (shown in gray). The field star spectrum has been normalized to that of Cen X-4 just blueward of the $^{12}$CO bandhead at 2.29 $\mu$m.}
\end{figure}

\clearpage

\begin{figure}
\figurenum{15}
\epsscale{1.0}
\plotone{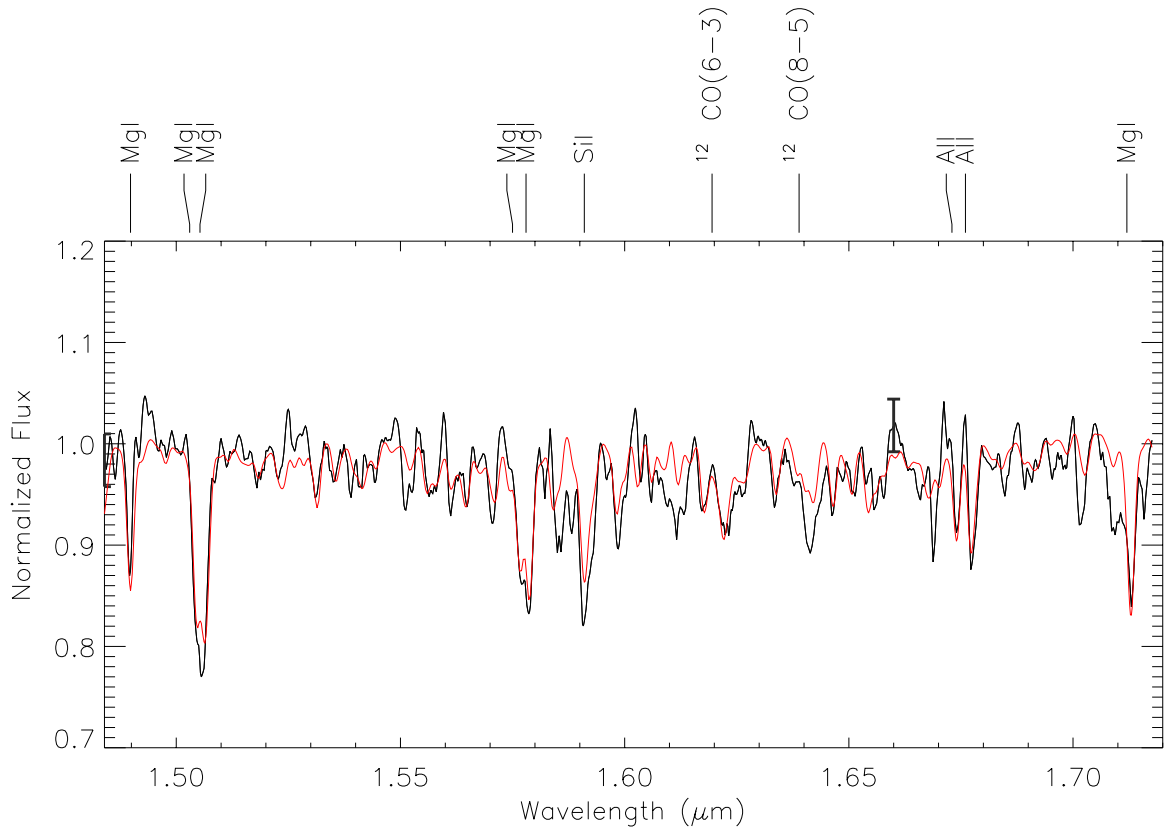}
\caption{The normalized H-band spectrum of CenX-4 is shown in black. Overplotted in red is a K7V field star, normalized and scaled to $f=0.94$. Representative error bars are shown in the figure}
\end{figure}

\clearpage

\begin{figure}
\figurenum{16}
\epsscale{1.0}
\plotone{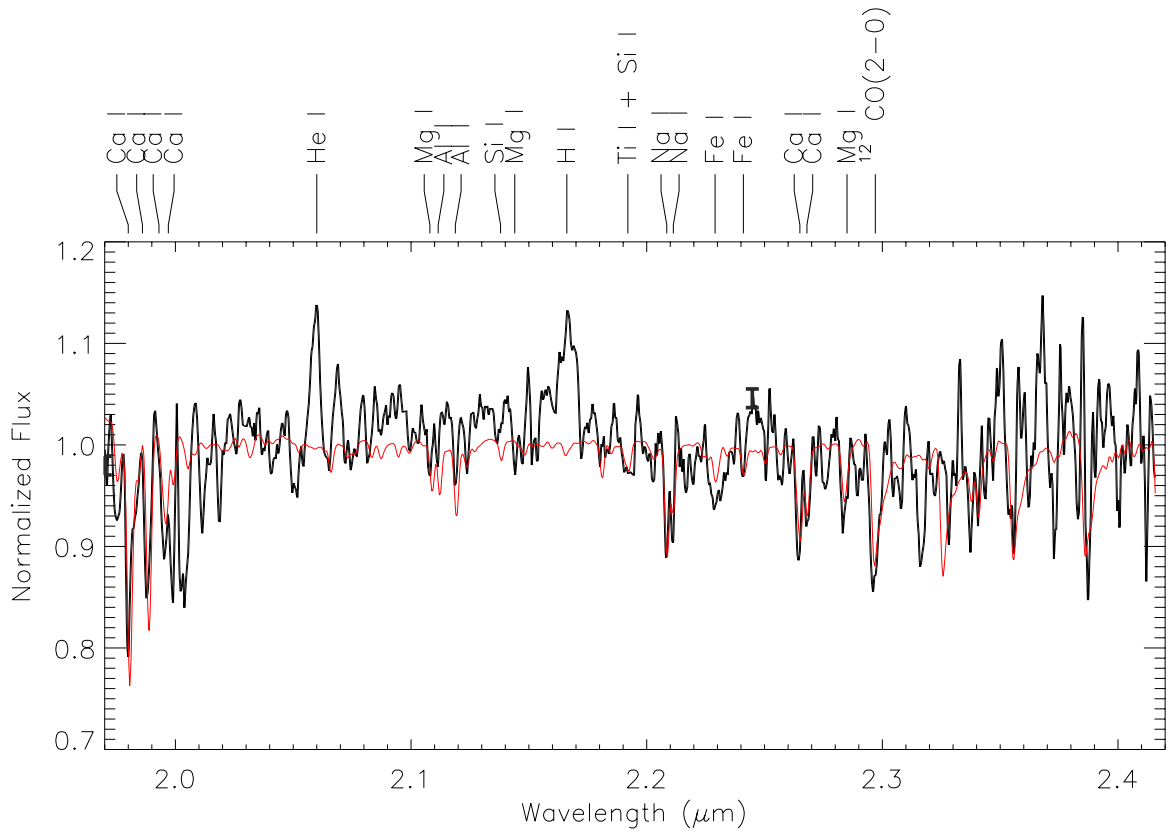}
\caption{The normalized K-band spectrum of CenX-4 is shown in black. Overplotted in red is the spectrum of an M0 V field star, normalized and scaled to $f=1.00$ with representative error bars.}
\end{figure}

\clearpage

\begin{figure}
\figurenum{17}
\epsscale{1.0}
\plotone{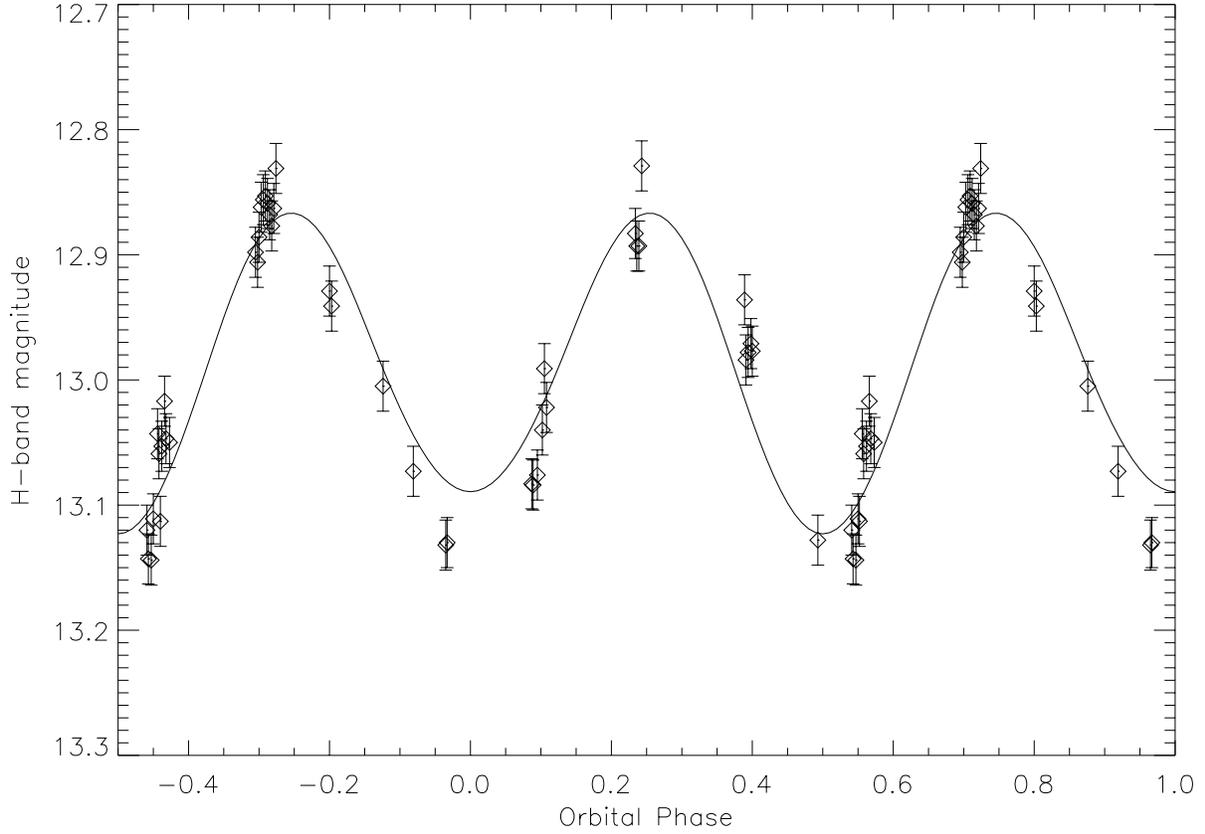}
\caption{The best fit  H-band light curve of V404 Cyg using the orbital ephemeris of \cite{casares94}. The data points (diamond) and the equally weighted error bars for each point are obtained from \citet{sanwal1996}. The solid line shows the best fit to the ellipsoidal modulation including 2\% dilution from the accretion disk for an inclination of 67$^\circ$ and q = 16.67 }
\end{figure}

\clearpage

\begin{figure}
\figurenum{18}
\epsscale{1.0}
\plotone{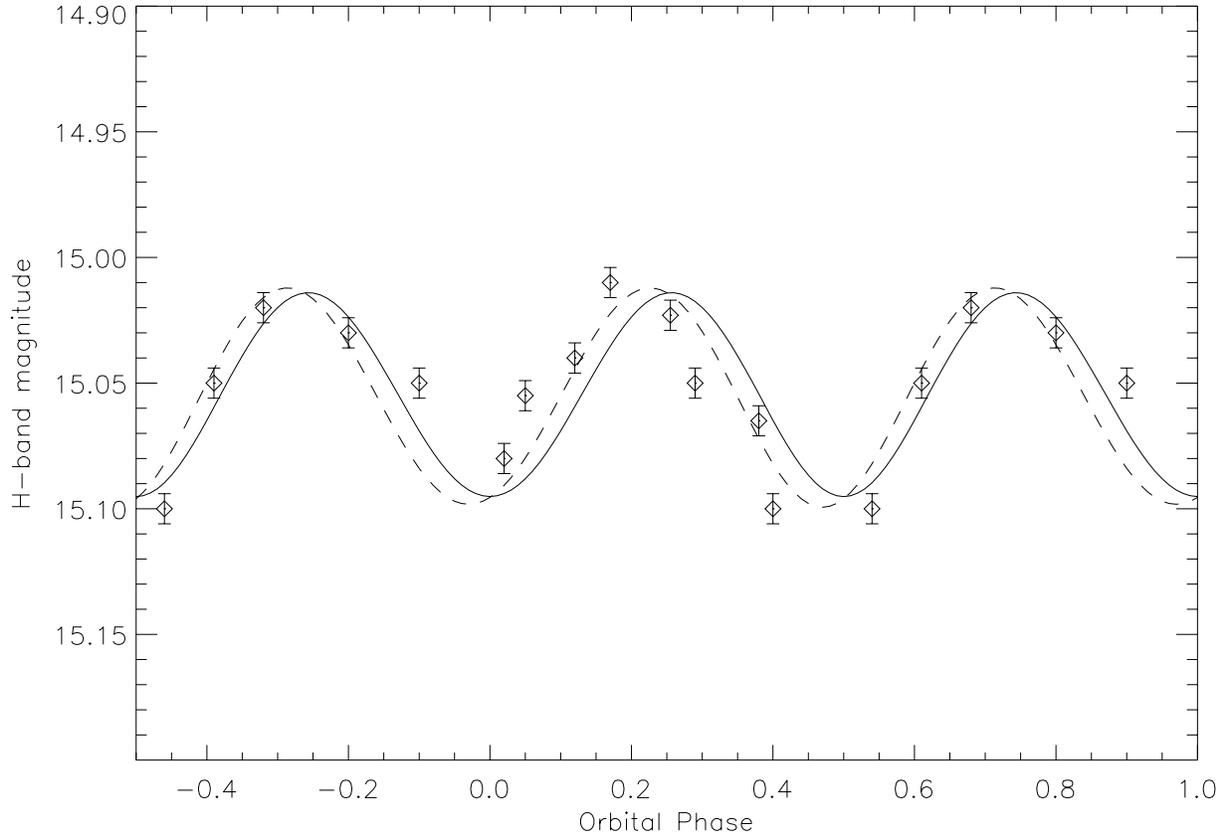}
\caption{The best fit H-band light curve of Cen X-4 with zero phase-offset and an inclination of 35$^\circ$ is shown in solid  while the dashed line represents the best fit for an inclination of 36$^\circ$ and a phase-offset of .03. The data points (diamond) and the equally weighted error bars for each point are obtained from \citet{shahbaz1993}. Both include 6\% dilution from the accretion disk and are plotted for q = 5.00.}  
\end{figure}

\end{document}